\documentclass[prd,aps,preprintnumbers,showpacs,nofootinbib,superscriptaddress]{revtex4}

\usepackage{amssymb,amsmath,graphicx,bm}

\DeclareMathOperator{\tr}{Tr}

\usepackage[usenames,dvipsnames]{color}
\usepackage[normalem]{ulem} 
\renewcommand\sout{\bgroup \color[rgb]{0.55,0.00,0.99} \ULdepth=-.5ex \ULset}
\newcommand{\xB}{x_{\scriptscriptstyle B}}
\newcommand{\sT}{{\scriptscriptstyle T}}
\renewcommand{\d}{\mathrm{d}}
\def\slash#1{\setbox0=\hbox{$#1$}               
        \dimen0=\wd0                            
        \setbox1=\hbox{/} \dimen1=\wd1          
        \ifdim\dimen0>\dimen1                   
        \rlap{\hbox to \dimen0{\hfil/\hfil}}    
        #1                                      
        \else
        \rlap{\hbox to \dimen1{\hfil$#1$\hfil}} 
        /                                       
        \fi}                                    %

\begin{document}


\title{Asymmetries in Heavy Quark Pair and Dijet Production at an EIC}

\author{Dani\"el Boer}
\email{d.boer@rug.nl}
\affiliation{ {Van Swinderen Institute for Particle Physics and Gravity, University of Groningen, Nijenborgh 4, 9747 AG Groningen, The Netherlands}}

\author{Piet J. Mulders}
\email{mulders@few.vu.nl}
\affiliation{Nikhef and Department of Physics and Astronomy, VU University
 Amsterdam, De Boelelaan 1081, NL-1081 HV Amsterdam,
The Netherlands}

\author{Cristian Pisano}
\email{cristian.pisano@unipv.it}
\affiliation{Dipartimento di Fisica, Universit\`a di Pavia, via Bassi 6, I-27100 Pavia, Italy}
\affiliation{INFN Sezione di Pavia, via Bassi 6, I-27100 Pavia, Italy}

\author{Jian Zhou}
\email{jzhou@sdu.edu.cn} \affiliation{\normalsize\it School of physics, $\&$ Key Laboratory of
Particle Physics and Particle Irradiation (MOE), Shandong University, Jinan, Shandong 250100,China}
\affiliation{Nikhef and Department of Physics and Astronomy, VU University
 Amsterdam, De Boelelaan 1081, NL-1081 HV Amsterdam,
The Netherlands}

\begin{abstract}
Asymmetries in heavy quark pair and dijet production in electron-proton collisions allow studies of
gluon TMDs in close analogy to studies of quark TMDs in semi-inclusive DIS. Here we present expressions
for azimuthal asymmetries for both unpolarized and transversely polarized proton cases and consider
the maximal asymmetries allowed. The latter are found to be rather sizeable, except in certain 
kinematic limits which are pointed out. In addition, we consider the small-$x$ limit and expectations 
from a McLerran-Venugopalan model for unpolarized and linearly polarized gluons and from
a perturbative, large transverse momentum calculation for the T-odd gluon TMDs. 
Comparison to related observables at RHIC and LHC is expected to provide valuable information 
about the process dependence of the gluon TMDs. In particular this will offer the possibility of a sign 
change test of the gluon Sivers TMD and two other T-odd gluon TMDs. This provides additional motivation 
for studies of azimuthal asymmetries in heavy quark pair and dijet production at a future Electron-Ion Collider.
\end{abstract}

\date{\today}

\maketitle

\section{Introduction}
Heavy quark pair and dijet production in electron-proton collisions offer direct probes of the
gluons inside the proton. They allow to map out detailed information about the momentum
distribution of the gluons, including the transverse momentum
dependence~\cite{Boer:2010zf,Pisano:2013cya}. This is of particular interest in the case of
transversely polarized protons, which are known to exhibit large left-right asymmetries. Such
asymmetries have not yet been studied in heavy quark pair or dijet production in electron-proton
collisions, but form a prominent part of the physics case for an Electron-Ion Collider (EIC)
proposed in the U.S.~\cite{Boer:2011fh,Accardi:2012qut}. Especially the study of the gluon
Sivers effect is of interest here. The Sivers effect~\cite{Sivers:1989cc} refers to the transverse
momentum distribution of unpolarized quarks and gluons inside a transversely polarized proton,
where the transverse momentum forms a $\sin\phi$ distribution around the transverse spin direction.

For transversely polarized protons there is a strong analogy between quark and gluon distributions.
For both there is the Sivers effect already mentioned, but  there is also a parallel between the
helicity flip distributions of transversely polarized quarks and of linearly polarized gluons
inside transversely polarized protons, for which there are two each. Even though the latter
distributions have different chirality and T behavior, their transverse momentum structures are
such that they lead to the same azimuthal asymmetries in certain processes. This parallel we will
discuss explicitly for the quark asymmetries in semi-inclusive Deep Inelastic Scattering (SIDIS)
process ($e\, p \to e' \, h\, X$) and the gluon asymmetries in open heavy quark pair production
($e\, p \to e' Q\, \overline{Q}\, X$).

The transverse momentum dependent distribution functions (TMDs) of gluons inside transversely
polarized protons are also of interest because of their process dependence and their associated
small $x$ behavior. TMDs are inherently process dependent, expressed through their gauge link
dependence. For the processes under consideration here, the gauge links are exclusively future
pointing, and can be related directly to the gluon TMDs appearing at LHC in Higgs or heavy scalar
quarkonium production (see section \ref{sec:links}). This in principle allows to cross check the results obtained at LHC and EIC
for unpolarized protons and perhaps for transversely polarized protons too in case a polarized
fixed target experiment, called AFTER@LHC~\cite{Brodsky:2012vg,Lansberg:2014myg}, will be performed
in the future at LHC. The gluon TMDs appearing in these processes correspond to the
Weisz\"acker-Williams (WW) distributions at small
$x$~\cite{Dominguez:2011wm,Dominguez:2010xd}. Recently it was pointed out in
Ref.~\cite{Boer:2015pni} that unlike the dipole gluon TMDs (which have one future and one past pointing gauge link),
the WW gluon TMDs for a transversely polarized proton 
are suppressed with respect to the unpolarized gluon TMD by a factor of $x$. This implies that the transverse spin
asymmetries in heavy quark pair and dijet production in electron-proton collisions will become
suppressed in the small-$x$ limit. A test of this at EIC would be very interesting, as the EIC can
probe both small and large $x$ regions. At RHIC heavy quarkonia in the forward region are not
accessible, but other processes can be studied that in the small-$x$ limit probe the dipole
distributions that are not suppressed with respect to the unpolarized gluon TMD towards small $x$.
Such processes may also be studied very precisely at large $x$ at
AFTER@LHC~\cite{Schafer:2013wca,Kanazawa:2015fia,Ma:2015vpt}. For a further discussion we
refer to Ref.~\cite{Boer:2015ika}. Obtaining a consistent picture among the various processes at
small and large $x$ is important for testing our understanding of transverse spin effects and the
TMD formalism in general.

In this paper we will restrict to the WW-type functions that appear in the heavy quark pair and
dijet production at an EIC. The outline of the paper is as follows. In Section~\ref{sec:def}
we provide the definition of gluon TMDs in terms of QCD operators and discuss some phenomenological
models which are relevant in the small-$x$ region. Details of the calculations of the cross section
for heavy quark pair production in DIS are discussed in Section~\ref{sec:HQ}, together with the
expressions for the azimuthal asymmetries and their upper bounds. In section \ref{sec:links} we comment 
on the process dependence of the gluon TMDs involved. Our results for
dijet production are given in Section~\ref{sec:dijet}. Summary and conclusions are presented in
Section~\ref{sec:concl}.

\section{Gluon TMD definitions and small-$x$ expressions}
\label{sec:def}

Transverse momentum dependent distribution functions (TMDs) of gluons inside a proton with four-momentum $P$ and spin vector $S$ are defined through a matrix element of a correlator of the gluon field strengths $F^{\mu \nu}(0)$ and $F^{\nu \sigma}(\xi)$, evaluated at fixed light-front (LF) time $\xi^+ =\xi{\cdot}n=0$, where $n$ is a lightlike vector conjugate to $P$.  Explicitly, the correlator is given by~\cite{Mulders:2000sh}
\begin{eqnarray}
\label{GluonCorr}
 {\Gamma}_g^{\mu\nu}(x,\bm p_\sT )
& = &  \frac{n_\rho\,n_\sigma}{(P{\cdot}n)^2}
{\int}\frac{d(\xi{\cdot}P)\,d^2\xi_\sT}{(2\pi)^3}\ e^{ip\cdot\xi}\, \langle P,
S|\,\tr\big[\,F^{\mu\rho}(0)\, U_{[0,\xi]} F^{\nu\sigma}(\xi)\,U^\prime_{[\xi, {0}]}\,\big] \,|P, S
\rangle\,\big\rfloor_{\text{LF}}\,,
\end{eqnarray}
where the gluon momentum $p$ is decomposed as
$p = x\,P + p_\sT + p^- n$,  while $U_{[0,\xi]}$ and $U^\prime_{[0,\xi]}$ are process dependent gauge links that ensure gauge invariance. The longitudinal and transverse component of the proton spin are defined
through the Sudakov decomposition
\begin{equation}
S^\mu = \frac{S_L}{M_p}\, \bigg (  P^\mu -  \frac{M_p^2}{P\cdot n}\, n^\mu\bigg )  + S_\sT^\mu\,,
\end{equation}
with $S_L^2 + \bm S_\sT^2= 1$ and $M_p$ being the proton mass. For an unpolarized ($U$) and transversely polarized ($T$) proton, the following parametrizations for the correlator in terms of gluon TMDs will be employed:
\begin{eqnarray}
 {\Gamma}_U^{\mu\nu}(x,\bm p_\sT ) & = & \frac{x}{2}\,\bigg \{-g_\sT^{\mu\nu}\,f_1^g (x,\bm p_\sT^2) +\bigg(\frac{p_\sT^\mu p_\sT^\nu}{M_p^2}\,
    {+}\,g_\sT^{\mu\nu}\frac{\bm p_\sT^2}{2M_p^2}\bigg) \,h_1^{\perp\,g} (x,\bm p_\sT^2) \bigg \} , \nonumber \\
 {\Gamma}_T^{\mu\nu}(x,\bm p_\sT ) &  = & \frac{x}{2}\,\bigg \{g^{\mu\nu}_\sT\,
    \frac{ \epsilon^{\rho\sigma}_\sT p_{\sT \rho}\, S_{\sT\sigma}}{M_p}\, f_{1T}^{\perp\,g}(x, \bm p_\sT^2) + i \epsilon_\sT^{\mu\nu}\,
    \frac{p_\sT \cdot S_\sT}{M_h}\, g_{1T}^{g}(x, \bm p_\sT^2) \nonumber \\
    & & \,  + \,  \frac{p_{\sT \rho}\,\epsilon_\sT^{\rho \{ \mu}p_\sT^{\nu \}}}{2M_p^2}\,\frac{p_\sT\cdot S_\sT }{M_p} \, h_{1 T}^{\perp\,g}(x, \bm p_\sT^2)\,- \,\frac{p_{\sT \rho} \epsilon_\sT^{\rho \{ \mu}S_\sT^{\nu \}}\, + \,
      S_{\sT\rho} \epsilon_\sT^{\rho \{ \mu } p_\sT^{\nu \}}}{4M_p} \, h_{1T}^{g}(x, \bm p_\sT^2) \,\,\bigg \}\,,
\label{eq:Phipar}
\end{eqnarray}
where we have  introduced the symmetric and antisymmetric transverse projectors $g^{\mu\nu}_{\sT} = g^{\mu\nu} - P^{\mu}n^{\nu}/P{\cdot}n-n^{\mu}P^{\nu}/P{\cdot}n$ and $\epsilon_\sT^{\mu\nu}  = \epsilon_\sT^{\alpha\beta\mu\nu} P_\alpha n_\beta/P\cdot n$, with $\epsilon_\sT^{1 2} = +1$, respectively.

The symmetric part of the correlator $\Gamma_T$, i.e.\ $(\Gamma_T^{\mu\nu}+\Gamma_T^{\nu\mu})/2$,
is parametrized by three gluon TMDs that are all T-odd.
The naming scheme used here for the gluon TMDs is based on
the analogy with the quarks~\cite{Meissner:2007rx}, but it should be emphasized
 that the quark and gluon TMDs with the same name are not necessarily linked by evolution
 and can have quite different properties under symmetry operations.
 The $h$ functions for quarks are chiral-odd and do not mix with the chiral-even $h$ functions of the gluons.
 Below we will make use of the combination
\begin{equation}
h_1^g \equiv h_{1T}^g +\frac{\bm p_\sT^2}{2 M_p^2}\,  h_{1T}^{\perp\,g}\,,
\label{eq:h1}
\end{equation}
which, despite its name, is not related to the well-known transversity distribution $h_1^q$ for
quarks that has no gluonic analogue. Although both $h_1^q$ and $h_1^g$ denote helicity
flip distributions, the quark distribution is chiral-odd, T-even, and survives after transverse
momentum integration, while the gluon distribution is chiral-even, T-odd, and vanishes upon
integration over transverse momentum. Both distributions require another helicity flip somewhere
else in the process, which will require consideration of different types of processes. But as we will
show, the angular dependences in such distinct processes can be the same, which in our view justifies the
naming scheme.

\begin{figure}[t]
\includegraphics[angle=0,width=0.45\textwidth]{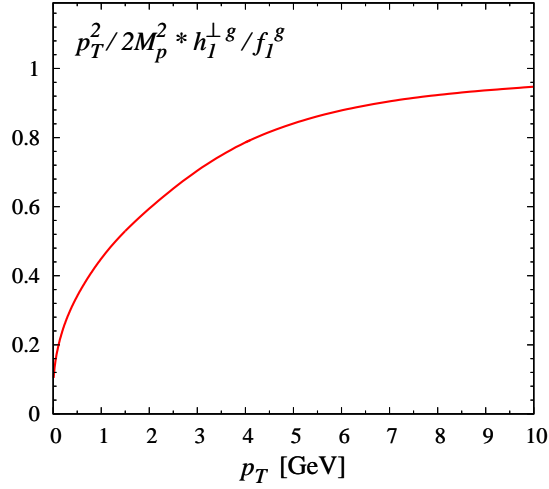}
\caption{Ratio of the linearly polarized and unpolarized WW-type gluon TMDs in the MV model.}
\label{fig:bound-mv}
\end{figure}

In order to estimate the size of the asymmetries in the small-$x$ limit, we discuss
here some aspects of the small-$x$ behavior of the various gluon TMDs, starting with the
linearly polarized gluon distribution. It has been shown that 
both the dipole and WW type linearly polarized gluon distributions can be studied in
the small-$x$ formalism as they both have the same ${\rm ln}\,1/x$ enhancement in the small $x$-region as the unpolarized gluon distributions. It is thus possible to compute them using a saturation model. For example, in the McLerran-Venugopalan (MV)
model~\cite{McLerran:1993ni}, the WW type linearly polarized gluon distribution is given
by~\cite{Metz:2011wb,Dominguez:2011br}
\begin{eqnarray}
x h^{\perp g}_{1}(x, \bm p_\sT^2)=\frac{S_\perp}{2\pi^3 \alpha_s} \frac{M_p^2}{2 \bm p_\sT^2}
\frac{N_c^2-1 }{ N_c} \int_0^\infty d r_\perp \frac{J_2 \left (|\bm p_\sT| r_\perp \right ) } { r_\perp
{\rm ln} \frac{1}{ r_\perp^2 \Lambda_{QCD}^2}} \left ( 1-e^{ - \frac{r_\perp^2 Q_s^2}{4}} \right
)\,,
\end{eqnarray}
where the saturation scale is defined as $Q_s^2= \alpha_s N_c \mu_A{\rm ln} \frac{1}{ r_\perp^2
\Lambda_{QCD}^2}$ and $S_\perp$ is the transverse area of the nucleon/nucleus. In the MV model,
there is a relation $\mu_A S_\perp=\alpha_s 2\pi A$, where $A$ is the number of nucleons inside a
nucleus, which is 1 in our case. Although the MV model is argued to work better for a large
nucleus, its application to a nucleon target turns out to be quite successful
phenomenologically~\cite{Albacete:2010sy,Lappi:2013zma}. For numerical estimations a regularized MV
model expression is required for the small-$x$ WW type gluon TMDs~\cite{Albacete:2003iq,
Enberg:2005cb,Lappi:2013zma},
\begin{eqnarray}
x h^{\perp g}_{1}(x, \bm p_\sT^2)=\frac{ (N_c^2-1 )\alpha_s}{\pi^2} \frac{M_p^2}{2 \bm p_\sT^2}
\int_0^\infty d r_\perp \frac{J_2 \left (|\bm p_\sT| r_\perp \right ) } { r_\perp Q_{s0}^2 {\rm ln}
\left [ \frac{1}{ r_\perp^2 \Lambda^2_{QCD}}+e\right ]} \left \{ 1-e^{ - \frac{r_\perp^2
Q_{s0}^2}{4}{\rm ln} \left [ \frac{1}{r_\perp^2 \Lambda^2_{QCD} }+e \right ]} \right \}\,,
\end{eqnarray}
where $Q_{s0}^2$ is a starting saturation scale to be taken from fits to experimental data. The ratio between
linearly polarized and unpolarized gluon distributions in the MV model is then given by
\begin{equation}
\frac{\bm p_\sT^2 }{2M_p^2}\,\frac{h_1^{\perp\,g}(x, \bm p_\sT^2)}{f_1^g(x, \bm p_\sT^2)} = \frac{
\int_0^\infty \frac{\d r_\perp}{r_\perp}\,  J_2 \left (| \bm p_\sT| r_\perp \right ) {{\rm ln}^{-1}
\left [ \frac{1}{r_\perp^2 \Lambda^2_{QCD}}+e\right ]}\, \left \{ 1-e^{ - \frac{r_\perp^2
Q_{s0}^2}{4}{\rm ln} \left [ \frac{1}{r^2_\perp \Lambda^2_{QCD} }+e \right ]} \right
\}}{\int_0^\infty \frac{\d r_\perp}{r_\perp}\,  J_0 \left (| \bm p_\sT| r_\perp \right ) \, \left
\{ 1-e^{ - \frac{r_\perp^2 Q_{s0}^2}{4}{\rm ln} \left [ \frac{1}{r_\perp^2 \Lambda^2_{QCD} }+e
\right ]} \right \}}\,.
\end{equation}
Numerically we find that $Q_{s0}^2$ should be  larger than about 1.1 times $C_A
\Lambda_{QCD}^2/(4C_F)$, in order for the ratio to stay below 1  for all $p_\sT$ values (this is
contrary to the claim in \cite{Dominguez:2011br}). Alternatively one could consider replacing $e$
in the numerator by $e_c \cdot e$, and take $e_c$ to be sufficiently larger than 1. In
Fig.~\ref{fig:bound-mv} we show the ratio for the case $Q_{s0}^2= C_A/C_F \times 0.35$ GeV$^2$ at
$x=0.01$ and $\Lambda_{QCD} = 0.2$ GeV. Here the value $0.35$ GeV$^2$ is based on the 
GBW model fits to HERA data \cite{GolecBiernat:1998js}. If one uses instead the value $0.104$ GeV$^2$ 
in combination with $\Lambda_{QCD} = 0.241$ GeV, as in~\cite{Lappi:2013zma}, the resulting function 
would violate the positivity bound in a restricted region of intermediate transverse momentum. 
As can be seen from the figure, unlike the dipole case for which the ratio
is saturated for all $p_\sT$ \cite{Metz:2011wb}, the ratio in this WW case is not saturated for
$p_\sT$ values below 10 GeV, but still sizeable except for very small $p_\sT$ values. Sizeable
asymmetries may thus also be expected at EIC at small but not too small $x$.

Now we turn to discuss the behavior of the WW type T-odd gluon TMDs in the small $ x$ limit.
 At large $p_\sT$, three T-odd WW type gluon TMDs inside a transversely polarized target
can be perturbatively computed~\cite{Boer:2015pni} following the standard collinear twist-3
approach~\cite{Ji:2006ub,Zhou:2008fb}. As mentioned in the introduction, these calculations  have
suggested that the WW gluon TMDs are suppressed by a factor of $x$ as compared to the corresponding
dipole type ones.  However, in Ref.~\cite{Boer:2015pni}, only the asymptotic small-$x$ behavior of
the WW gluon TMDs is discussed in the context of the collinear twist-3 approach without giving the
full expressions. Below we present the complete results for these gluon distributions, which are
relevant at moderately small $x$. One notices that such sub-asymptotic behavior can not be
studied in the MV model, as in general, only the leading logarithm contribution is taken into
account in saturation models. For the WW type gluon Sivers function, one has
\begin{eqnarray}
  f_{1T}^{\perp\,g} (x,\bm p_\sT^2)&=& \frac{C_1M_p}{p_T^4} \left \{ \int_{x}^{1} \frac{\d z}{z}
\sum_{q+\bar{q}}
 \left \{T_{F,q} (z,z) \frac { 1+(1-\xi)^2}{\xi} -T_{F,q}(z,z-x)  \frac{2-\xi}{\xi}
\right \}  \right .\ \nonumber \\
&&- 16 \pi M_p \int_x^1 \frac{d z}{z^2} \left \{
 \frac{(\xi^2-\xi+1)^2}{\xi(1-\xi)_+} \left [ N(z,z)
 -  N(z,0)\right ] \right .\
 + \frac{1+\xi^2}{2\xi(1-\xi)_+} N(x,z)
 \nonumber \\ & & - \!\!\!\!\!\! \left .\ \left .\
  \frac{1+(1-\xi)^2}{2\xi(1-\xi)_+} N(z,z-x)
 -\frac{\xi^2+(1-\xi)^2}{2\xi(1-\xi)_+}N(x,x-z)
  \right \} \right \}\,,
\label{fsivers}
\end{eqnarray}
where $\xi=x/z$ and $C_1=\frac{N_c}{2} \frac{\alpha_s}{2\pi^2}$. Here $T_{F,q}(z,z)$ is the
collinear twist-3 quark-gluon correlation function, commonly known as the Qiu-Sterman
function~\cite{Qiu:1991pp}, while $N(z_1,z_2)$ is a tri-gluon correlation function in which three
gluons are in a symmetric color singlet state~\cite{Ji:1992eu,Beppu:2010qn,Schafer:2013opa}. The
tri-gluon correlation function has the symmetry properties: $ N(x_1,x_2) = N(x_2, x_1) , \
N(x_1,x_2) = -N(-x_1,-x_2)$, which below will be used to simplify the expressions. 
Terms proportional to $ \delta(1-\xi) [N(z,z)-N(z,0)]$ are ignored, as they are suppressed in the limit $x \to 0$.
The large-$p_T$ expression of the gluon TMD $h_{1T}^{\,g} (x,p_T^2)$ takes a similar form,
\begin{eqnarray}
h_{1T}^{\,g} (x, \bm p_T^2)& =& C_1 \frac{M_p}{p_T^4} 2 \left \{ \int_{x}^{1}  \frac{\d z}{z}
\sum_{q+\bar{q}} \left \{T_{F,q} (z,z) \frac { 2-2\xi }{\xi} -T_{F,q}(z,z-x) \frac{2-\xi}{\xi}
\right \}, \right .\
 \nonumber \\
&&- 16 \pi M_p \int_x^1 \frac{d z}{z^2} \left \{
  \frac{1-\xi}{\xi}  N(z,z)
 -  \frac{\xi^2+(1-\xi)^2}{\xi(1-\xi)_+} N(z,0) \right .\
 + \frac{1-\xi}{2\xi} N(x,z)
 \nonumber \\ & & - \!\!\!\!\! \left .\ \left .\
  \frac{1+(1-\xi)^2}{2\xi(1-\xi)_+} N(z,z-x)
 -\frac{1}{2\xi(1-\xi)_+}N(x,x-z)
  \right \} \right \}\, ,
\end{eqnarray}
where again the term proportional to $\delta(1-\xi)$ is ignored. Finally, the gluon TMD $h_{1T}^{\perp\, g} (x,p_T^2)$ is
given by
\begin{eqnarray}
 h_{1T}^{\perp\, g} (x,p_T^2) =
 C_1  \frac{M_p}{p_T^4} \frac{2M_p^2}{p_T^2} \int_{x}^{1} \frac{dz}{z}\left \{ \sum_{q+\bar{q}}
T_{F,q} (z,z)+ T_{G}^{(+)}(z,z) \right \} \frac {4-4\xi}{\xi} \,,
\end{eqnarray}
where the tri-gluon correlation function $T_{G}^{(+)}(z,z) $ is defined as $T_G^{(+)}(z,z)\equiv-8 \pi M_p\left [ N(z,z)- N(z,0) \right ]/z$.
  With these results, we can study the small-$x$ asymptotic behavior of the WW  T-odd gluon TMDs. First
of all, as observed in \cite{Boer:2015pni}, there is a complete cancelation between the soft gluon
pole contributions and the hard gluon pole contributions in the leading power of $1/x$ for the
gluon distributions $ f_{1T}^{\perp \, g} (x,p_T^2)$ and $h_{1T}^{\,g} (x,p_T^2)$ in the small-$x$
limit.  The leading contribution to the gluon TMD $h_{1T}^{\perp\, g} (x,p_T^2)$ also vanishes by
invoking the Burkardt sum rule~\cite{Burkardt:2003yg,Lorce:2015lna},
\begin{eqnarray}
 \int_0^1 dz \left \{ \sum_{q+\bar{q}}
T_{F,q} (z,z)+ T_{G}^{(+)}(z,z) \right \}=0\,,
\end{eqnarray}
which is stable under QCD evolution~\cite{Zhou:2015lxa}. However, beyond the leading logarithm
${\rm ln}\, 1/x$ approximation, there is no exact cancellation among the various contributions. In
the small-$x$ limit, the subleading contributions to the gluon TMDs are given by
\begin{eqnarray}
 f_{1T}^{\perp \, g} (x,\bm p_T^2) &\approx& -\frac{C_1M_p}{p_T^4}  \int_{x}^{1} \frac{dz}{z}
\left \{ \sum_{q+\bar{q}} T_{F,q} (z,z)+2T_{G}^{(+)}(z,z) \right \}  \,,\\
h_{1T}^{\,g} (x, \bm p_T^2) &\approx& -\frac{C_1M_p}{p_T^4}2  \int_{x}^{1} \frac{dz}{z} \left \{
\sum_{q+\bar{q}} T_{F,q} (z,z)+2T_{G}^{(+)}(z,z) \right \}\,,
\\
h_{1T}^{\perp \, g} (x, \bm p_T^2) &\approx&-
  \frac{ C_1 M_p}{p_T^4} 4\frac{2M_p^2}{p_T^2}\int_{x}^{1} \frac{dz}{z}\left \{ \sum_{q+\bar{q}}
T_{F,q} (z,z)+ T_{G}^{(+)}(z,z) \right \} .
\end{eqnarray}
 We emphasize that the calculation presented here
is based on the large-$p_\sT$ DGLAP-type formalism. Unlike the T-odd dipole type  gluon
TMDs~\cite{Boer:2015pni,Zhou:2013gsa}, these T-odd WW type gluon TMDs cannot be treated in the
small-$x$ formalism due to the lack of the leading logarithmic contribution, at least not in a
conventional way~\cite{Bartels:1995iu}. Nevertheless, we will use these compact
results as guidance for modelling the WW gluon TMDs at small $x$ for all $p_\sT$. 
For example, if we assume the Burkardt sum
rule were satisfied at each value of $z$, i.e.\ $ \sum_{q+\bar{q}} T_{F,q} (z,z)+
T_{G}^{(+)}(z,z)=0$, one obtains the simple result
\begin{eqnarray}
 f_{1T}^{\perp \, g} (x, \bm p_T^2)=\frac{1}{2} h_{1T}^{\,g} (x, \bm p_T^2), \  \  \ h_{1T}^{\perp \, g}
 (x, \bm p_T^2)=0 \quad \Rightarrow\ h_1^g(x, \bm p_T^2)= 2 f_{1T}^{\perp \, g} (x, \bm p_T^2) \qquad ({\rm Model}\ {\rm I}).
\end{eqnarray}
If, on the other hand, we assume that the tri-gluon correlation were zero at some low initial scale and is purely
dynamically generated by the Qiu-Sterman function $T_{F,q}$, one would have (note that $p_\sT^2=-\bm p_\sT^2$)
\begin{eqnarray}
 f_{1T}^{\perp \, g} (x, \bm p_T^2)=\frac{1}{2} h_{1T}^{\,g} (x, \bm p_T^2)=
 \frac{p_T^2}{8M_p^2} h_{1T}^{\perp \, g}  (x, \bm p_T^2) \quad \Rightarrow\ h_1^g(x, \bm p_T^2)= - 2 f_{1T}^{\perp \, g} (x, \bm p_T^2) \qquad ({\rm Model}\ {\rm II}).
\end{eqnarray}
Finally, if for some reason it turns out that $\int_{x}^{1} dz T_{G}^{(+)}(z,z)/z \gg
\sum_{q+\bar{q}} \int_{x}^{1} dz T_{F,q} (z,z)/z$, one obtains:
\begin{eqnarray}
 f_{1T}^{\perp\, g } (x, \bm p_T^2)=\frac{1}{2} h_{1T}^{\,g} (x, \bm p_T^2)=
 \frac{p_T^2}{4M_p^2} h_{1T}^{\perp\, g } (x, \bm p_T^2) \quad \Rightarrow\ h_1^g(x, \bm p_T^2)= 0\qquad ({\rm Model}\ {\rm III}).
\end{eqnarray}
These models will be used below to obtain simple estimates for some ratio of asymmetries. All
three models differ from the leading logarithmic result for the T-odd dipole functions
$f_{1T}^{\perp \, g}=\frac{1}{2}h_{1T}^{\,g}= \frac{p_\sT^2}{2M_p^2} h_{1T}^{\perp\, g}=h_1^g$.

The T-odd gluon TMDs have to satisfy the following positivity bounds~\cite{Mulders:2000sh}
\begin{align}
\frac{\vert \bm p_\sT \vert }{M_p}\, \vert f_{1T}^{\perp \,g}(x,\bm p_\sT^2) \vert & \le   f_1^g(x,\bm p_\sT^2)\,,\nonumber \\
\frac{\vert \bm p_\sT \vert }{M_p}\, \vert h_{1}^g(x,\bm p_\sT^2) \vert & \le   f_1^g(x,\bm p_\sT^2)\,,\nonumber \\
\frac{\vert \bm p_\sT \vert^3}{2 M_p^3}\, \vert h_{1T}^{\perp \,g}(x,\bm p_\sT^2) \vert & \le   f_1^g(x,\bm p_\sT^2)\,.
\end{align}
In none of the three models we presented, one can satisfy the bounds for all three functions at
the same time. This is in contrast to the dipole model for which all bounds can be saturated simultaneously.

It should be emphasized that for small $x$ the maximal values are not expected for the WW-type functions that are probed at EIC in heavy quark pair or dijet production. 
As mentioned, the WW T-odd gluon TMDs are suppressed by a factor of $x$ with respect to the WW unpolarized gluon TMD.
The WW T-even gluon TMD $h_1^{\perp \, g}$ is not suppressed by a factor of $x$, but by a logarithm of $p_\sT/Q_s$ in
the region $p_\sT \ll Q_s$ \cite{Metz:2011wb}, whereas for $p_\sT \gg Q_s$ it was found to saturate the bound. This is compatible with the sizeable but not maximal effects we obtained in the MV model, Fig.\ \ref{fig:bound-mv}. This is why below we will show results for the MV model, rather than for $h_1^{\perp \, g}$ that saturates the positivity bound for all $p_\sT$.

\section{Heavy quark pair production}
\label{sec:HQ}
\begin{figure}[t]
\begin{center}
 \includegraphics[angle=0,width=0.48\textwidth]{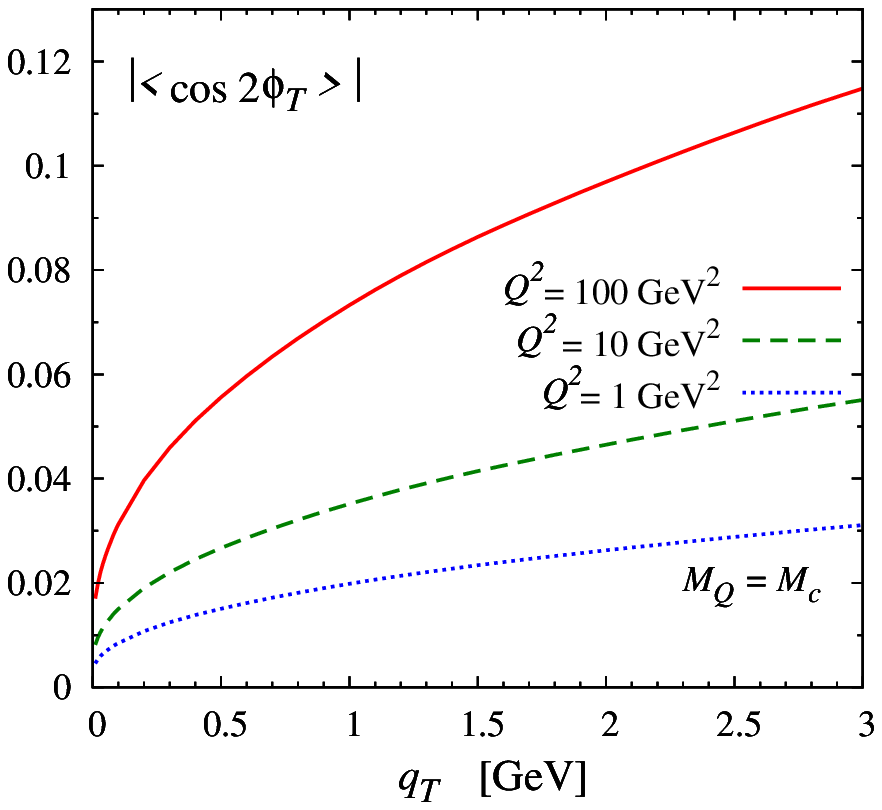}
 \includegraphics[angle=0,width=0.48\textwidth]{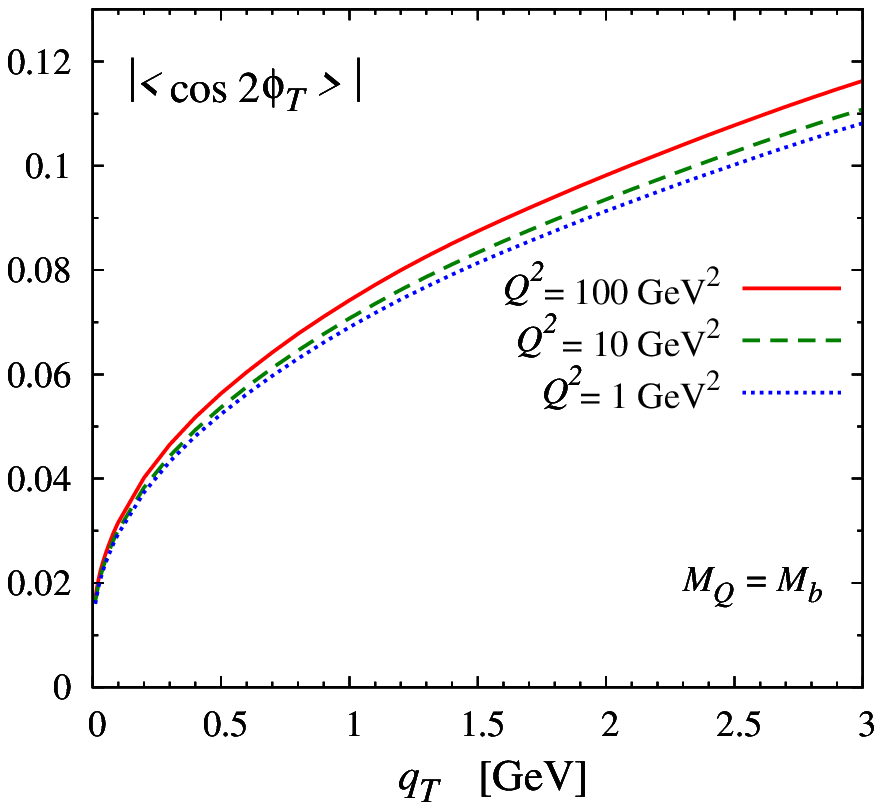}
 \caption{Estimates  of the asymmetry $\vert \langle\cos 2 \phi_\sT \rangle \vert $  in the MV model, as a function of $q_\sT \equiv \vert \boldsymbol q_\sT\vert$, calculated at $ \vert \boldsymbol K_\perp \vert = 6 $ GeV,  $z=0.5$,  $y=0.1$ and at different values of  $Q^2$,  for charm (left panel) and bottom (right panel) production in the process $e p\to e^\prime Q \overline{Q} X$.}
\label{figRp}
\end{center}
\end{figure}

The calculation of the cross section for the process
\begin{equation}
e(\ell) + p(P,S) \to e(\ell^{\prime}) + {Q}(K_1) +  {\overline Q}(K_2)+ X\,,
\end{equation}
where the proton is polarized with polarization vector $S$ and the other particles are unpolarized, proceeds along the same lines of Ref.~\cite{Pisano:2013cya}, to which we refer for details. The heavy quark-antiquark pair in the final state is almost back-to-back in the plane perpendicular to the direction of the incoming proton and the photon exchanged in the reaction, that we take as $\hat z$-axis. Therefore one has $\vert q_\sT \vert \ll \vert K_\perp\vert $, where $q_\sT \equiv K_{1\perp} + K_{2\perp}$ and $K_\perp \equiv (K_{1\perp} -K_{2\perp})/2$, with $K_{i \perp}$ ($i=1,2$) being the transverse momenta of the heavy quark and antiquark, satisfying the relation $K_{i\perp}^2 = -\bm K_{i\perp}^2$. This is often referred to as the back-to-back correlation limit. In a reference frame in which azimuthal angles are measured w.r.t.\ the lepton plane ($\phi_{\ell}=\phi_{\ell^\prime}=0$),
denoting by $\phi_S$, $\phi_\sT$ and $\phi_\perp$ the azimuthal angles of the three-vectors $\bm S_\sT$,  $\bm q_\sT$ and  $\bm K_\perp$, respectively, the cross section can be written as
\begin{equation}
\frac{\d\sigma}
{\d y_1\,\d y_2\,\d y\,\d\xB\,\d^2\bm{q}_{\sT} \d^2\bm{K}_{\perp}} \equiv \d\sigma (\phi_S, \phi_\sT,\phi_\perp) =    \d\sigma^U(\phi_\sT,\phi_\perp)  +  \d\sigma^T (\phi_S, \phi_\sT,\phi_\perp)  \,,
\label{eq:cs}
\end{equation}
where $y_i$ are the rapidities of the quarks, $y$ is the inelasticity variable, and  $x_B=Q^2/(2 P\cdot q)$, with $Q^2= -q^2= -(\ell-\ell^\prime)^2$, is the Bjorken-$x$ variable.
At leading order in perturbative QCD, we find
\begin{align}
\d\sigma^U
  & =  {{\cal N}}\, \bigg [ A_0^U + A_1^U \cos \phi_\perp  + A_2^U \cos 2 \phi_\perp +  B_0^U \cos 2 \phi_\sT +  B_1^U \cos( 2\phi_\sT-\phi_\perp) \nonumber \\
& \qquad \quad+ B_2^U \cos 2 (\phi_\sT-\phi_\perp) + B_3^U \cos( 2\phi_\sT-3\phi_\perp)  + B_4^U\cos( 2\phi_\sT-4\phi_\perp)  \bigg ]  \delta(1-z_1-z_2) \,,
\label{eq:csU}
\end{align}
and
\begin{align}
\d\sigma^T
  & =  {{\cal N}\,\vert \bm S_\sT\vert}\, \bigg \{ \sin(\phi_S -\phi_\sT) \bigg [ A_0^T + A_1^T \cos \phi_\perp  + A_2^T \cos 2 \phi_\perp \bigg ] + \cos(\phi_S-\phi_\sT) \bigg [ B_0^T \sin 2 \phi_\sT \nonumber \\
& \quad + \, B_1^T \sin(2\phi_\sT-\phi_\perp) + B_2^T \sin 2 (\phi_\sT-\phi_\perp) + B_3^T \sin( 2\phi_\sT-3\phi_\perp)  + B_4^T\sin(2\phi_\sT- 4\phi_\perp)  \bigg ] \nonumber\\
& \quad + \bigg [B_0^{\prime\, T} \sin(\phi_S+\phi_\sT) + B_1^{\prime \,T} \sin(\phi_S + \phi_T-\phi_\perp) + B_2^{\prime\,T } \sin(\phi_S+\phi_\sT-2\phi_\perp) \nonumber \\
& \qquad \quad  +  B_3^{\prime\, T} \sin(\phi_S+\phi_\sT-3\phi_\perp) + B_4^{\prime\, T} \sin(\phi_S+\phi_\sT-4 \phi_\perp) \bigg ] \bigg \}\,\delta(1-z_1-z_2)\, ,
\label{eq:csT}
\end{align}
where $z_i = P \cdot K_i/P \cdot q$, with $q\equiv \ell-\ell^\prime$.
The normalization factor  $\cal N$ is given by
\begin{equation}
{\cal N} =   \frac{\alpha^2\alpha_s}{ \pi s M_\perp^2}\, \frac{1}{ \xB y^2} \,,
\end{equation}
with $s=(\ell+P)^2$, $M_\perp= \sqrt{M_Q^2 + \bm K_\perp^2}$, where $M_Q$ is the (anti)quark mass.
The explicit expressions of the terms $A_l^{U/T}$ in Eqs.~(\ref{eq:csU}) and (\ref{eq:csT}), with $l=0,1,2$, are
given by
\begin{equation}
A_{l}^U
 =e_Q^2 \,T_R\,{\cal A}_l^{e g\to e Q  {\overline Q} }\, f_{1}^{g} (x,\bm{q}_{\sT}^2) \, ,\qquad
A_{l}^T = A_{l}^U\, \frac{f_{1T}^{\perp\, g} (x,\bm{q}_{\sT}^2)\  }{f_{1}^{g} (x,\bm{q}_{\sT}^2)}\,,
\end{equation}
where $T_R=1/2$. We note that the three ratios $A_l^T/A_l^U$ are all equal to each other,  i.e.\ to the ratio of the gluon Sivers function $f_{1T}^{\perp\,g}$ and the unpolarized gluon distribution $f_1^g$.
The hard scattering functions ${\cal A}_l^{e g\to e Q  {\overline Q}}$ describe the interaction of an unpolarized gluon with a photon which can be in different polarization states~\cite{Pisano:2013cya},
\begin{align}
{\cal A}_0^{e g\to e Q {\overline  Q}} & =  [1+(1-y)^2]\,{\cal A}_{U+L}^{\gamma^*g\to Q  {\overline Q}} -y^2 \,{\cal A}_{L}^{\gamma^*g \to Q  {\overline Q}}\,,\nonumber \\
{\cal A}_{1}^{e g\to e Q {\overline  Q}} & = (2-y) \sqrt{1-y}\,{\cal A}_{I}^{\gamma^*g \to Q  {\overline Q}}\,,\nonumber \\
 {\cal A}_{2}^{e g\to e Q {\overline Q}} & =  2 (1-y)\, {\cal A}_{T}^{\gamma^*g \to Q {\overline Q}}\,,\nonumber \\
\label{eq:Agstar}
\end{align}
where the subscripts $U+L$, $L$, $I$, $T$ refer to the specific polarization of the photon~\cite{Pisano:2013cya,Brodkorb:1994de}.  Introducing the notation $z\equiv z_1$, $z_2=1-z$, $Q^2=-q^2$,  we have
\begin{align}
{\cal A}_{U+L}^{\gamma^* g\to Q  {\overline Q}} & =  \frac{1}{D^3} -\frac{z(1-z)}{D^3}\,\left \{ 2 -4\,\frac{M_Q^2}{M_\perp^2} + 4\,\frac{M_Q^4}{M_\perp^4} - \left [4z(1-z)\left (2 -3\,\frac{M_Q^2}{M_\perp^2}\right ) +2\,\frac{M_Q^2}{M_\perp^2} \right ]\frac{Q^2}{M_\perp^2} \,\right . \nonumber \\
& \qquad\ \left . -\, z(1-z)[1-2 z(1-z)]\frac{Q^4}{M_\perp^4} \right \}\,, \label{eq:AQQb2}\\
{\cal A}_{L}^{\gamma^* g\to Q  {\overline Q}} & =  8\, \frac{z^2(1-z)^2}{D^3} \left (1-\frac{M_Q^2}{M_\perp^2}\right )\frac{Q^2}{M_\perp^2}\,, \\
{\cal A}_{I}^{\gamma^* g\to Q  {\overline Q}} & =  4 \,\sqrt{1-\frac{M_Q^2}{M_\perp^2}} \,
\frac{z(1-z)(1-2 z)}{D^3}\,\frac{Q}{M_\perp}
\bigg [1- z (1-z) \frac{Q^2}{M_\perp^2} -2 \frac{M_Q^2}{M_\perp^2} \bigg ]\,,
\label{eq:AQQbphil}\\
{\cal A}_{T}^{\gamma^* g\to Q  {\overline Q}} & =  4\,\frac{z(1-z)}{D^3}\,\bigg (1-\frac{M_Q^2}{M_\perp^2} \bigg)
\bigg [ z (1-z) \frac{Q^2}{M_\perp^2} + \frac{M_Q^2}{M_\perp^2} \bigg ]\, ,
\label{eq:AQ}
\end{align}
with
\begin{equation}
D\equiv D \left (z, \frac{Q^2}{M_\perp^2} \right ) = 1 + z (1-z) \frac{Q^2}{M_\perp^2}~.
\label{eq:Den}
\end{equation}
Analogously, the $B_m^U$, $B_m^T$ and  $B_m^{\prime\,T} $, with $m=0,1,2,3,4$, describe the scattering of a linearly polarized gluon with a photon. They can be written as
\begin{equation}
B_{m}^U  = e_Q^2 \,T_R\,{\cal B}_m^{e g\to e Q  {\overline Q}} \, \frac{ \bm q_\sT^2  }{M_p^2} \,h_{1}^{\perp\, g} (x,\bm{q}_{\sT}^2)\, ,\qquad B_{m}^T  = B_{m}^U\, \frac{\vert \bm q_\sT \vert }{M_p}\, \frac{h_{1T}^{\perp\, g} (x,\bm{q}_{\sT}^2)}{h_{1}^{\perp\, g} (x,\bm{q}_{\sT}^2)}\, , \qquad B^{\prime\,T}_{m} = B_{m}^U\,\frac{M_p}{\vert \bm q_\sT\vert }\,  \frac{h_{1T}^{g} (x,\bm{q}_{\sT}^2)}{ h_1^{\perp\, g} (x,\bm{q}_{\sT}^2)}\,, \nonumber
\end{equation}
where
\begin{align}
{\cal B}_2^{e g\to e Q {\overline Q}} & =  [1+(1-y)^2]\,{\cal B}_{U+L}^{\gamma^*g\to Q  {\overline Q}} -y^2 \,{\cal B}_{L}^{\gamma^*g \to Q {\overline Q}}\,,\nonumber \\
{\cal B}_{j}^{e g\to e Q {\overline Q}} & = (2-y) \sqrt{1-y}\,{\cal B}_{j I}^{\gamma^*g \to Q  {\overline Q}}\qquad (j =1,3) \,,\nonumber \\
 {\cal B}_{k}^{e g\to e Q {\overline Q}} & =  2 (1-y)\, {\cal B}_{k T}^{\gamma^*g \to Q  {\overline Q}}\qquad (k = 0,4)\,,
\label{eq:Agstar}
\end{align}
with
\begin{align}
{\cal B}_{U+L}^{\gamma^*g\to Q  {\overline Q}}  & = \frac{z(1-z)}{D^3}\, \bigg \{ \bigg [1 - 6 z (1-z) \bigg ]\,\frac{Q^2}{M_\perp^2}  - 2\, \frac{M_Q^2}{M_\perp^2} \bigg \}  \bigg (1-\frac{M_Q^2}{M_\perp^2} \bigg ) \,,\nonumber \\
{\cal B}_{L}^{\gamma^*g\to Q {\overline Q}}  & = -  4\,\frac{z^2(1-z)^2}{D^3}\, \bigg(1-\frac{M_Q^2}{M_\perp^2} \bigg )\frac{Q^2}{M_\perp^2}\,,\nonumber \\
{\cal B}_{1 I}^{\gamma^*g \to Q  {\overline Q}} & = 2 \,\frac{z(1-z)(1-2 z)}{D^3}\, \bigg [ z (1-z) \frac{Q^2}{ M_\perp^2} + \frac{M_Q^2}{M_\perp^2} \bigg ] \sqrt{1-\frac{M_Q^2}{M_\perp^2}}\frac{Q}{M_\perp} \,,\nonumber \\
{\cal B}_{3 I}^{\gamma^*g \to Q {\overline Q}} & = -2 \,\frac{z(1-z)(1-2 z)}{D^3}\,\bigg [ 1 -\frac{M_Q^2}{M_\perp^2} \bigg ]^{3/2}\frac{Q}{M_\perp}\,,\nonumber \\
{\cal B}_{0 T}^{\gamma^*g \to Q  {\overline Q}} & =  -\frac{z(1-z)}{D^3} \bigg [ z (1-z)\frac{Q^2}{ M_\perp^2} + \frac{M_Q^2}{M_\perp^2} \bigg]^2\,,\nonumber \\
{\cal B}_{4 T}^{\gamma^*g \to Q  {\overline Q}} & =  \frac{z(1-z)}{D^3}\,\bigg (1 - \frac{M_Q^2}{M_\perp^2} \bigg )^2~.
\end{align}
The expressions involving $h_1^{\perp\, g}$ have appeared before in Ref.~\cite{Pisano:2013cya}, but here a few redefinitions are made in order to display the similarity between the contributions of the various $h$ functions. In particular, the following observable,
 \begin{equation}
\vert \langle \cos 2 \phi_\sT \rangle \vert = \left| \frac{\int 
\d \phi_\perp\d \phi_\sT
\, \cos 2 \phi_\sT \, \d\sigma}{\int \d \phi_\perp \d \phi_\sT
\, \d\sigma}\right| = \frac{ \bm{q}_\sT^2\, | B_0^U|}{
2 \, A_0^U} = \frac{\bm q_\sT^2}{2 M^2}\,\frac{|h_1^{\perp\, g}(x,\bm p_\sT^2 )|}{ f_1^{g}(x,\bm p_\sT^2 )} \, \frac{\vert {\cal B}_0^{e g\to e Q  {\overline Q} } \vert}{{\cal A}_0^{e g\to e Q  {\overline Q}}}\,,
\end{equation}
has been defined, which could provide direct access to  $h_1^{\perp\, g}$. Using the earlier MV model expressions for
the ratio between linearly polarized and unpolarized TMDs, in Fig.~\ref{figRp} we show our estimates for $\vert \langle \cos2\phi_\sT\rangle\vert $ at  $z=0.5$, $y=0.1$, $\vert \bm K_\perp\vert $= 6 GeV and $Q^2 \ge 1$ GeV$^2$, for both charm and bottom production. For this and the subsequent numerical studies we take the charm mass $M_c= 1.3$ GeV and the bottom mass $M_b= 4.2$ GeV. We observe sizeable asymmetries in this MV model. Of course, the center of mass energy has to be sufficiently large for these curves to indeed be in the small-$x$ range. For an EIC the considered $\sqrt{s}$ ranges from about 20 to 150 GeV \cite{Boer:2011fh}, therefore $x_B = Q^2/(ys)<0.01$ for $y\geq 0.1$ requires $Q^2 < (10^{-3}-10^{-2}) s$. For $\sqrt{s}=60$ GeV one should thus not consider the $Q^2=100$ GeV$^2$ curve. We include the large $Q^2$ curves for a high energy EIC or for an analysis of HERA data which were taken at $\sqrt{s}=320$ GeV. For larger $y$ and $\vert \bm K_\perp\vert $ values the asymmetries decrease. For $y=1$ $\vert \langle \cos2\phi_\sT\rangle\vert $ vanishes identically.
\begin{figure}[t]
\begin{center}
 \includegraphics[angle=0,width=0.48\textwidth]{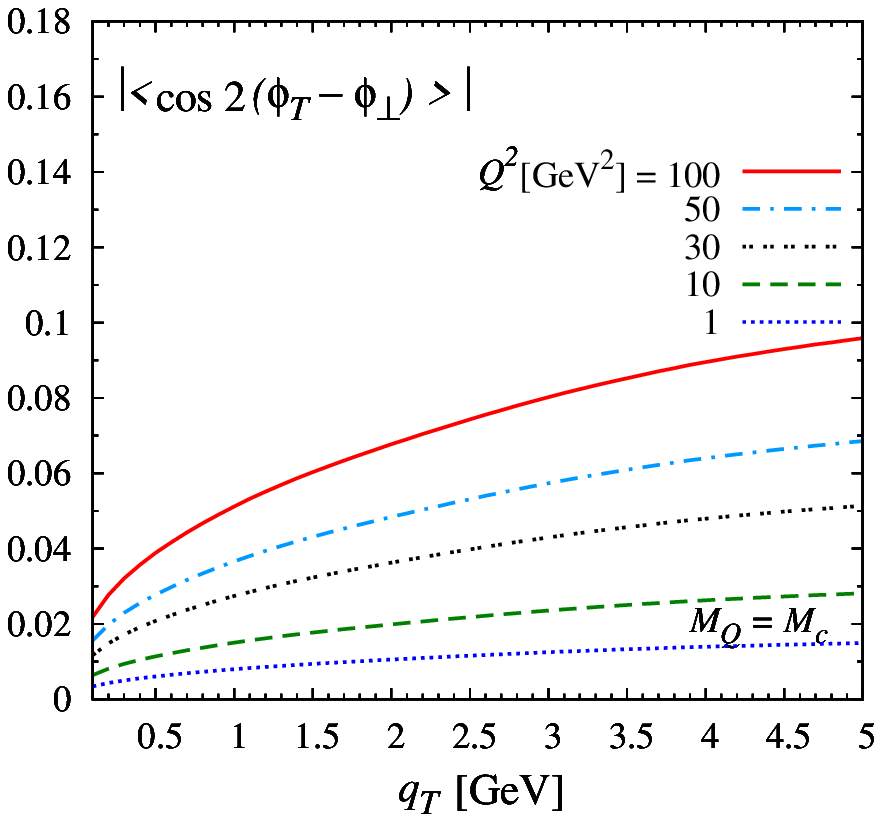}
 \includegraphics[angle=0,width=0.48\textwidth]{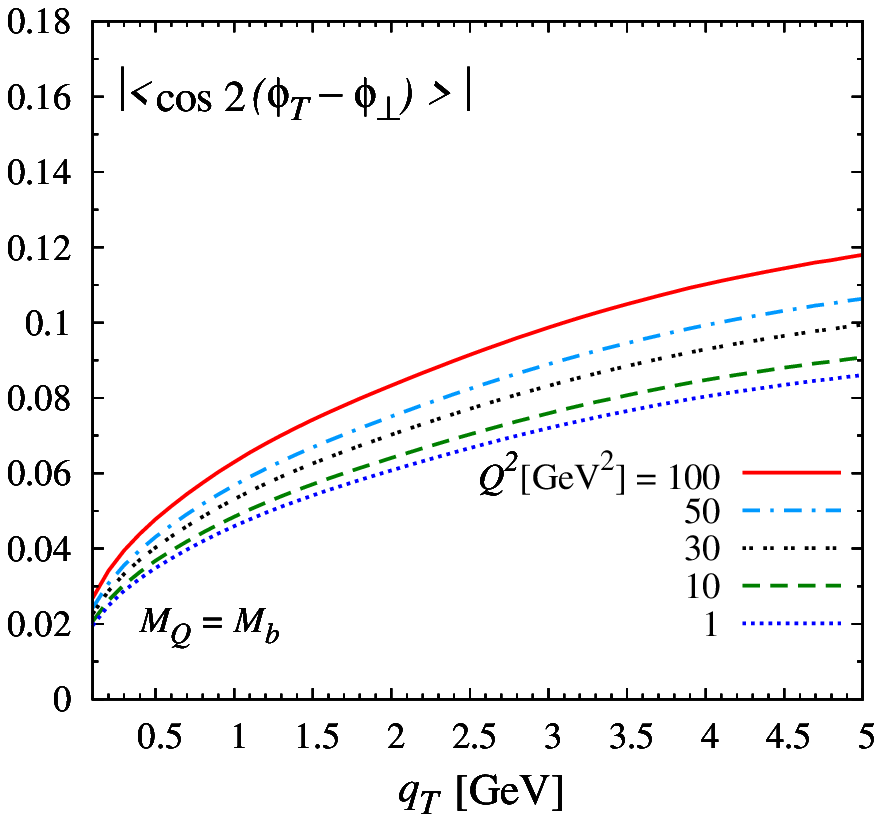}
\caption{Estimates  of the asymmetry $\vert \langle\cos 2 (\phi_\sT-\phi_\perp) \rangle \vert $  in the MV model, as a function of $q_\sT \equiv \vert \boldsymbol q_\sT\vert$, calculated at $ \vert \boldsymbol K_\perp \vert = 10 $ GeV,  $z=0.5$, $y=0.3$,  and different values of  $Q^2$, for charm (left panel) and  bottom (right panel) production in the process $e p\to e^\prime Q \overline{Q} X$.}
\label{figR}
\end{center}
\end{figure}

Analogously, in Fig.~\ref{figR} we provide our model estimates for the other asymmetry defined in Ref.~\cite{Pisano:2013cya},
\begin{equation}
\vert \langle \cos 2 (\phi_\sT-\phi_\perp) \rangle \vert = \left| \frac{\int 
\d \phi_\perp\d \phi_\sT
\, \cos 2 (\phi_\sT-\phi_\perp) \, \d\sigma}{\int \d \phi_\perp \d \phi_\sT
\, \d\sigma}\right| = \frac{ \bm{q}_\sT^2\, | B_2^U|}{
2 \, A_0^U} = \frac{\bm q_\sT^2}{2 M^2}\,\frac{|h_1^{\perp\, g}(x,\bm p_\sT^2 )|}{ f_1^{g}(x,\bm p_\sT^2 )} \, \frac{\vert {\cal B}_2^{e g\to e Q  {\overline Q}} \vert}{{\cal A}_0^{e g\to e Q  {\overline Q}}} \,,
\label{eq:R}
\end{equation} 
which is calculated at  $z=0.5$, $y=0.3$, $\vert \bm K_\perp\vert $= 10 GeV and $Q^2 \ge 1$ GeV$^2$. 
At large $q_\sT$ values the asymmetry is close to the maximal values allowed by the positivity bound. For larger $y$ and $\vert \bm K_\perp\vert $ values the asymmetries decrease again, but the asymmetry does not vanish at $y=1$.

If we introduce the combination $h_1^g$ of Eq.\ (\ref{eq:h1}), it turns out that the cross section in the transversely polarized proton case, when integrated over $\phi_\perp$, has only three independent azimuthal modulations:
$\sin(\phi_S-\phi_\sT)$, $\sin(\phi_S+\phi_\sT)$, and $\sin(\phi_S-3\phi_\sT)$. Each one of them is related to a different
T-odd gluon TMD. We observe that these angular structures and the TMDs they probe, are closely analogous to the case of 
quark asymmetries in SIDIS ($e\, p^\uparrow \to e' \, h\, X$), where the role of $\phi_T$ is played by $\phi_h$, cf.\ \cite{Boer:1997nt}. The same applies to the $\cos 2 (\phi_\sT-\phi_\perp)$ asymmetry considered before.

In order to single out separate $\phi_S$ dependent terms, we define the following azimuthal moments
\begin{align}
A_N^{W(\phi_S,\phi_\sT)} & \equiv 2\,\frac {\int \d \phi_\sT \,\d\phi_\perp\, W(\phi_S,\phi_\sT)\,\d\sigma_\sT(\phi_S,\,\phi_\sT,\,\phi_\perp)}{\int \d \phi_\sT \,\d\phi_\perp\,\d\sigma_U(\phi_\sT,\phi_\perp)} \nonumber \\
& =
2\, \frac {\int \d \phi_\sT \,\d\phi_\perp\, W(\phi_S,\phi_\sT)\,\left [\d\sigma(\phi_S,\,\phi_\sT,\,\phi_\perp) - \d\sigma(\phi_S + \pi,\,\phi_\sT,\,\phi_\perp)\right ]}{\int \d \phi_\sT \,\d\phi_\perp\, \left [\d\sigma(\phi_S,\,\phi_\sT,\,\phi_\perp) + \d\sigma(\phi_S + \pi,\,\phi_\sT,\,\phi_\perp)\right ]}  \,,
\label{eq:mom}
\end{align}
where the denominator is given by
\begin{align}
\int \d \phi_\sT \,\d\phi_\perp\,\d\sigma_U(\phi_\sT,\phi_\perp)  & \equiv \int \d \phi_\sT \,\d\phi_\perp\,\frac{\d\sigma_U(\phi_\sT,\phi_\perp)}{\d y_1\,\d y_2\,\d y\,\d\xB\,\d^2 \bm{q}_{\sT} \d^2\bm{K}_{\perp}} =   4 \pi\, \frac{\alpha^2\alpha_s}{ s M_\perp^2}\, \frac{1}{ \xB y^2}\, A_0^U\,,
\end{align}
and, since  $A_0^U = A_0^T\, f_1^g/f_{1T}^{\perp\,g}$, we obtain
\begin{align}
A_N^{\sin(\phi_S-\phi_\sT)} & = \frac{\vert \bm q_\sT\vert}{M_p}\, \frac{A_0^T}{A_0^U} =  \frac{\vert \bm q_\sT\vert}{M_p}\, \frac{f_{1T}^{\perp\,g}(x,\bm q_\sT^2)}{f_1^g(x,\bm q_\sT^2)}\,,
\label{eq:A1}\\
A_N^{\sin(\phi_S+\phi_\sT)}  & =  \frac{\vert \bm q_\sT\vert}{M_p}\,  \frac{B_0^{\prime \,T}}{A_0^U} =
\frac{2(1-y)\, {\cal B}_{0T}^{\gamma^*g \to Q  {\overline Q}} }{[1+(1-y)^2]{\cal A}^{\gamma^*g \to Q  {\overline Q}} _{U+L} -y^2 {\cal A}_L^{\gamma^*g \to Q  {\overline Q}} }
 \,\frac{\vert \bm q_\sT\vert}{M_p}\, \frac{h_{1}^{g}(x,\bm q_\sT^2)}{f_1^g(x,\bm q_\sT^2)}\,,
 \label{eq:A2}\\
A_N^{\sin(\phi_S-3\phi_\sT)}  & =   -  \frac{\vert \bm q_\sT\vert ^3}{M_p^3}\, \frac{B_0^T}{2 A_0^U} = - \frac{2(1-y)\, {\cal B}_{0T}^{\gamma^*g \to Q  {\overline Q}} }{[1+(1-y)^2]{\cal A}^{\gamma^*g \to Q  {\overline Q}} _{U+L} -y^2 {\cal A}_L^{\gamma^*g \to Q  {\overline Q}} }\, \frac{\vert \bm q_\sT\vert^3}{2 M_p^3}\, \frac{h_{1T}^{\perp \,g}(x,\bm q_\sT^2)}{f_1^g(x,\bm q_\sT^2)}\label{eq:A3}\,~.
\end{align}
Note that the asymmetries in Eqs.~(\ref{eq:A2}) and (\ref{eq:A3}) vanish in the limit $y\to 1$ when
the virtual photon is  longitudinally polarized, which for $s/Q^2 \to \infty$ corresponds
to the limit $x \to 0$. Based on the absence of the leading logarithmic term for the WW T-odd gluon TMDs, it is expected that also Eq.~(\ref{eq:A1}) vanishes in this limit, but it does not do so automatically through a kinematic suppression term as for the other two asymmetries.

We note that a measurement of the ratio
\begin{equation}
\frac{A_N^{\sin(\phi_S-3\phi_\sT)}}{A_N^{\sin(\phi_S+\phi_\sT)}} =- \frac{\bm q_\sT^2}{2 M_p^2}\, \frac{h_{1T}^{\perp\,g}(x,\bm q_\sT^2)}{h_{1}^{g}(x,\bm q_\sT^2)}
\end{equation}
would probe directly the relative magnitude of $h_{1T}^{\perp\,g}$ and $h_1^g$ without dependences on any of the other kinematic variables in the process. In the small-$x$ models discussed in the previous section, one would obtain for this ratio: 0 in model I, $-2$ in
model II, and $\infty$ in model III. These cases are all distinguishable from the dipole case, which would give 1. Note that this ratio need not be bounded between $-1$ and $+1$.

As mentioned before, the WW T-odd gluon TMDs are not expected to satisfy the positivity bounds, but the latter can be used to 
determine the maximum values of the above defined asymmetries, and thereby exclude less promising kinematic regions. It
can be easily seen that the Sivers asymmetry is bound to 1, while the asymmetries in
Eqs.~(\ref{eq:A2}) and (\ref{eq:A3}) have the same upper bound, which we denote by $A_N^W$.
The maximal values of $\vert A_N^W\vert $ for the latter are shown in Fig.~\ref{fig:ANW_y} for charm and bottom
 production as a function of $y$, at $z=1/2$, $Q^2= 1$, $10$ and $100$ GeV$^2$. These maximal values are attained 
 at some specific values of $ \vert  \bm K_\perp\vert$, which in this
case corresponds to the lowest $ \vert  \bm K_\perp\vert$ value considered. The charm and bottom
production upper bounds on $\vert A_N^W\vert $ as a function $ \vert  \bm K_\perp\vert$ are shown in
Fig.~\ref{fig:ANW}, again at $z=1/2$, $y=0.1$ and $Q^2= 1$, $10$ and $100$ GeV$^2$. The upper
bounds of these asymmetries are equal to the $R^\prime$ bounds of the weighted cross section
$\langle \cos2\phi_\sT\rangle$, which are presented in Fig.~2 of Ref.~\cite{Pisano:2013cya} for $y=0.01$.

Alternatively, one can define the azimuthal angles w.r.t.\ $\phi_\perp$ instead of $\phi_\ell$, which can then be integrated over. In this case, only the terms $A_0^T$, $B_2^T$ and $B_2^{\prime\,T}$  will contribute to the cross section in Eq.~(\ref{eq:csT}). After introducing again the combination in Eq.~(\ref{eq:h1}), one finds, in analogy to Eqs.~(\ref{eq:A1})-(\ref{eq:A3}),
\begin{align}
 A_N^{\sin(\phi_S^\perp-\phi_\sT^\perp)} & = A_N^{\sin(\phi_S-\phi_\sT)} =  \frac{\vert \bm q_\sT\vert}{M_p}\, \frac{A_0^T}{A_0^U} =  \frac{\vert \bm q_\sT\vert}{M_p}\, \frac{f_{1T}^{\perp\,g}(x,\bm q_\sT^2)}{f_1^g(x,\bm q_\sT^2)}\,, \\
A_N^{\sin(\phi_S^\perp+\phi_\sT^\perp)}  & =  \frac{\vert \bm q_\sT\vert}{M_p}\,  \frac{B^{\prime\,T}_2}{A_0^U} =
 \frac{[1+(1-y)^2 ]{\cal{B}}_{U+L}^{\gamma^*g \to Q  {\overline Q}}-y^2{\cal B}_L ^{\gamma^*g \to Q  {\overline Q}} }{[1+(1-y)^2]{\cal A}^{\gamma^*g \to Q  {\overline Q}} _{U+L} -y^2 {\cal A}_L^{\gamma^*g \to Q  {\overline Q}} }\,  \frac{\vert \bm q_\sT\vert}{M_p}\, \frac{h_{1}^{g}(x,\bm q_\sT^2)}{f_1^g(x,\bm q_\sT^2)}\,,
 \label{eq:A2b}\\
A_N^{\sin(\phi_S^\perp-3\phi_\sT^\perp)}  & =   \frac{\vert \bm q_\sT\vert ^3}{M_p^3}\, \frac{B_2^T}{2 A_0^U} =  \frac{[1+(1-y)^2 ]{\cal{B}}_{U+L}^{\gamma^*g \to Q  {\overline Q}}-y^2{\cal B}_L ^{\gamma^*g \to Q  {\overline Q}} }{[1+(1-y)^2]{\cal A}^{\gamma^*g \to Q  {\overline Q}} _{U+L} -y^2 {\cal A}_L^{\gamma^*g \to Q  {\overline Q}} }\, \frac{\vert \bm q_\sT\vert^3}{2 M_p^3}\, \frac{h_{1T}^{\perp \,g}(x,\bm q_\sT^2)}{f_1^g(x,\bm q_\sT^2)}\,~.
\label{eq:A3b}
\end{align}
As before, the asymmetries in Eqs.~(\ref{eq:A2b})-(\ref{eq:A3b}) have  the same upper bound
$A_N^{W_\perp}$, presented in Figs~\ref{fig:ANW_y_perp} and {\ref{fig:ANWp} for different values of
the kinematic variables $\vert \bm K_\perp\vert $, $y$, $z$, $Q^2$. This bound is equal to $R$, the
upper bound of $\langle \cos(2\phi_\sT-\phi_\perp) \rangle $ discussed in Fig.~1 of
Ref.~\cite{Pisano:2013cya} for $y=0.01$. From Fig.~\ref{fig:ANW_y_perp} one sees that in this case
the asymmetries do not need to vanish when $y \to 1$. This means that for small-$x$ studies the
asymmetries $A_N^{W_\perp}$ may be of more interest than $A_N^{W}$. Due to the zero crossing that
occurs in the maximal value of $|A_N^{W_\perp}|$, for charm production the asymmetry in the region of
$Q^2$ between 5 and 10 GeV$^2$ is expected to be small. For bottom production this would apply to
$Q^2$ closer to 100 GeV$^2$. Although these conclusions are based on the maximal asymmetries and
therefore robust, the exact location of the zero crossing in $y$ may be affected by higher order
corrections and thus may be a different function of $Q^2$ than obtained here. 

\begin{figure}[t]
\begin{center}
 \includegraphics[angle=0,width=0.48\textwidth]{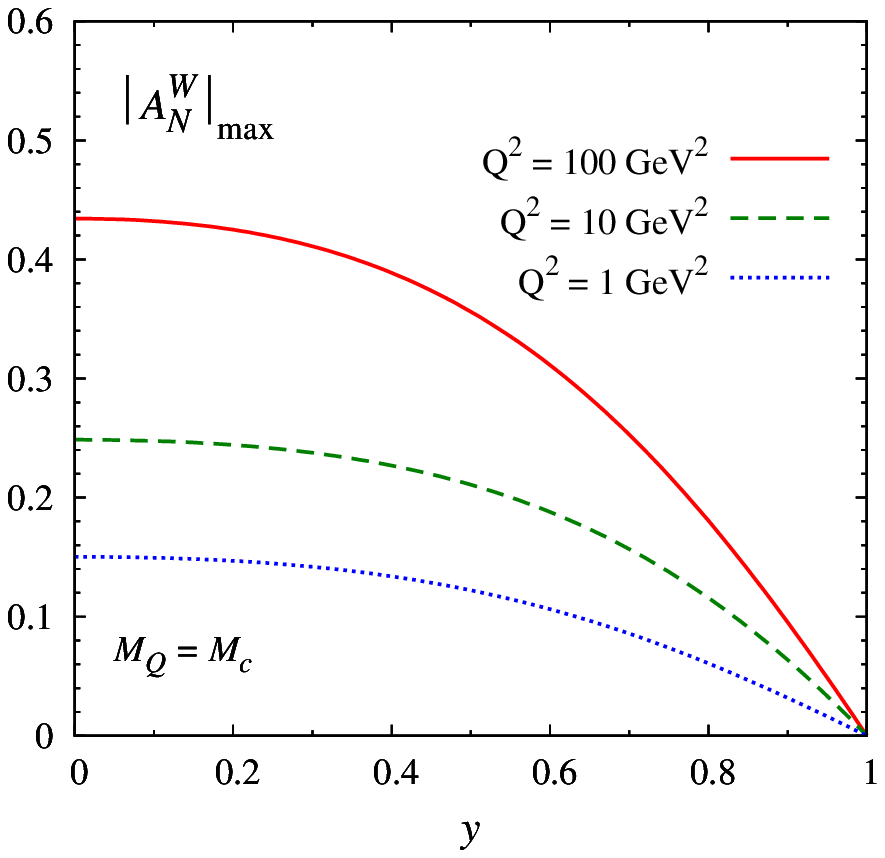}
 \includegraphics[angle=0,width=0.48\textwidth]{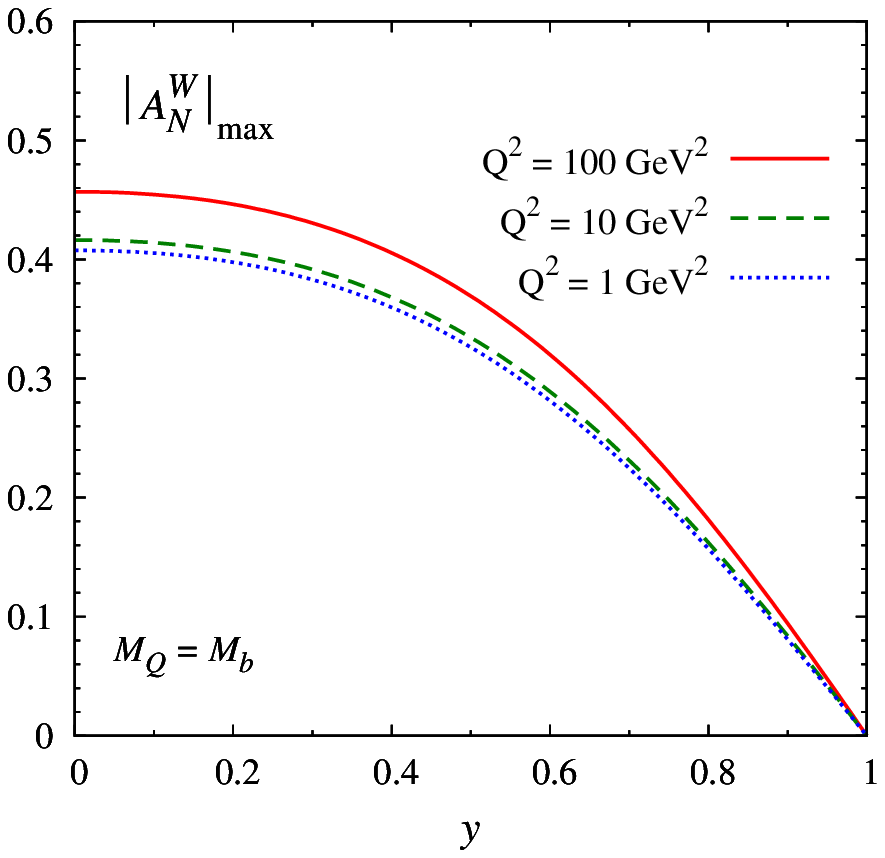}
 \caption{Upper bounds $\vert A_N^W\vert_{\rm max}$ on the asymmetries $A_N^{\sin(\phi_S+\phi_\sT)}$ and $A_N^{\sin(\phi_S-3\phi_\sT)}$  for charm (left panel) and bottom (right panel) production in process $e p\to e^\prime Q \overline{Q} X$. Estimates are given as a function of $y$, at different values of $K_\perp > 1$ GeV  (chosen in such a way that the asymmetries are maximal)  and $Q^2$, with $z=0.5$.}
\label{fig:ANW_y}
\end{center}
\end{figure}
\begin{figure*}[t]
\begin{center}
 \includegraphics[angle=0,width=0.48\textwidth]{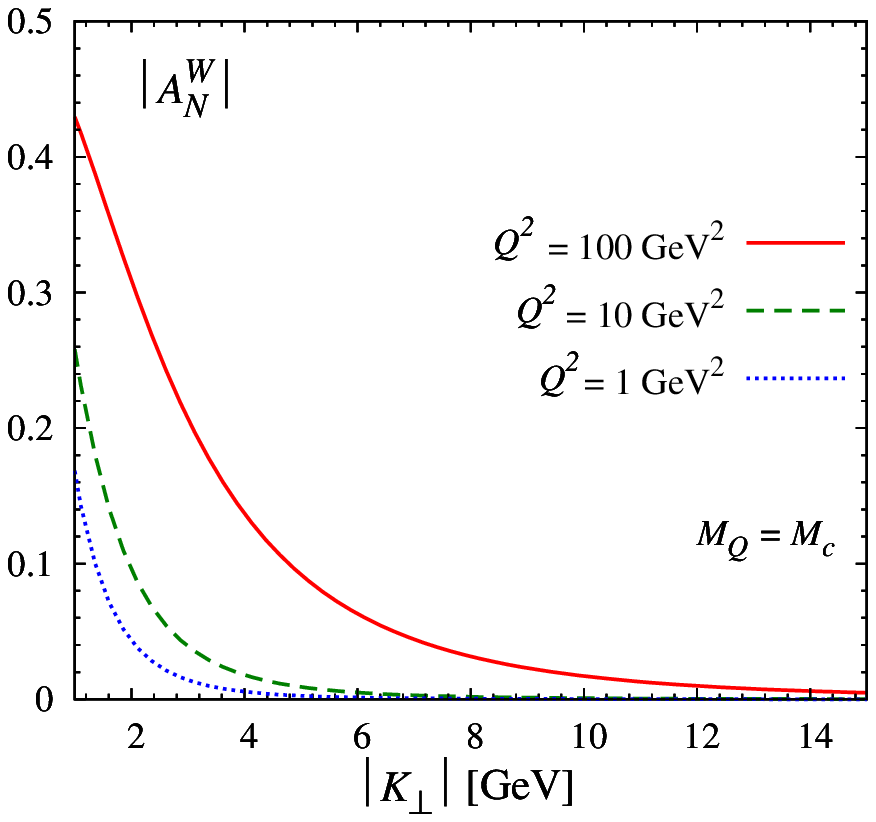}
 \includegraphics[angle=0,width=0.48\textwidth]{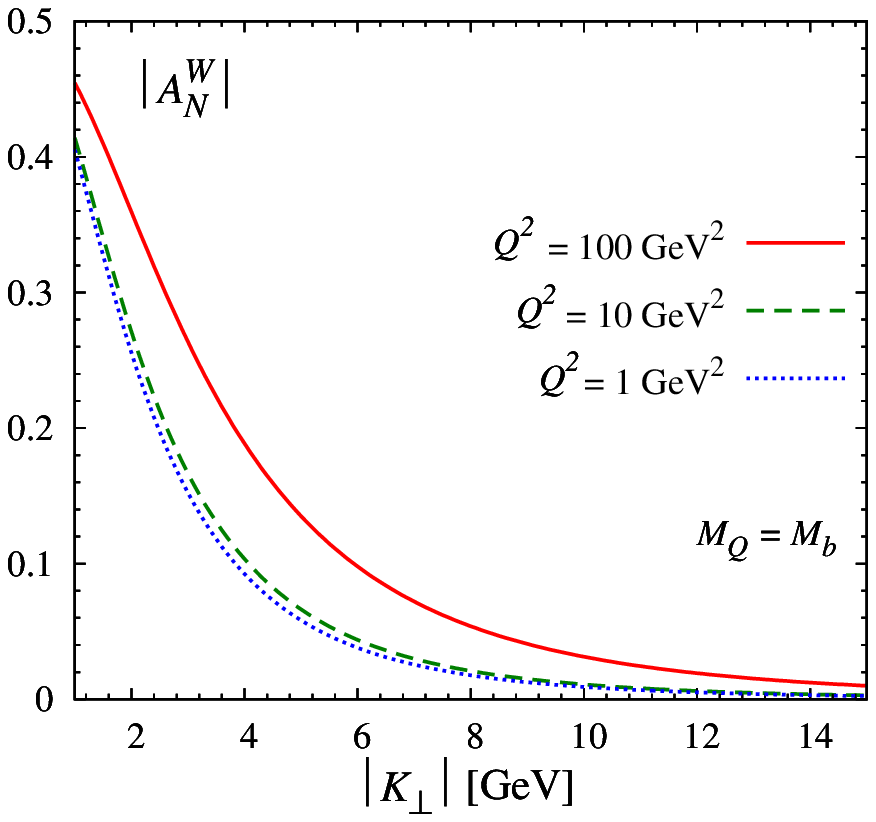}
 \caption{Estimates of $\vert A_N^W\vert $, with $W={\sin(\phi_S+\phi_\sT)}, \,{\sin(\phi_S-3\phi_\sT)}$, for charm
(left panel) and bottom (right panel) production in the process $e p\to e^\prime Q \overline{Q} X$, calculated at $z=0.5$,  $y=0.1$ and different values of  $Q^2$, as a function of $\vert \boldsymbol K_\perp\vert$ ($>$ 1 GeV).} \label{fig:ANW}
\end{center}
\end{figure*}
\begin{figure}[t]
\begin{center}
 \includegraphics[angle=0,width=0.48\textwidth]{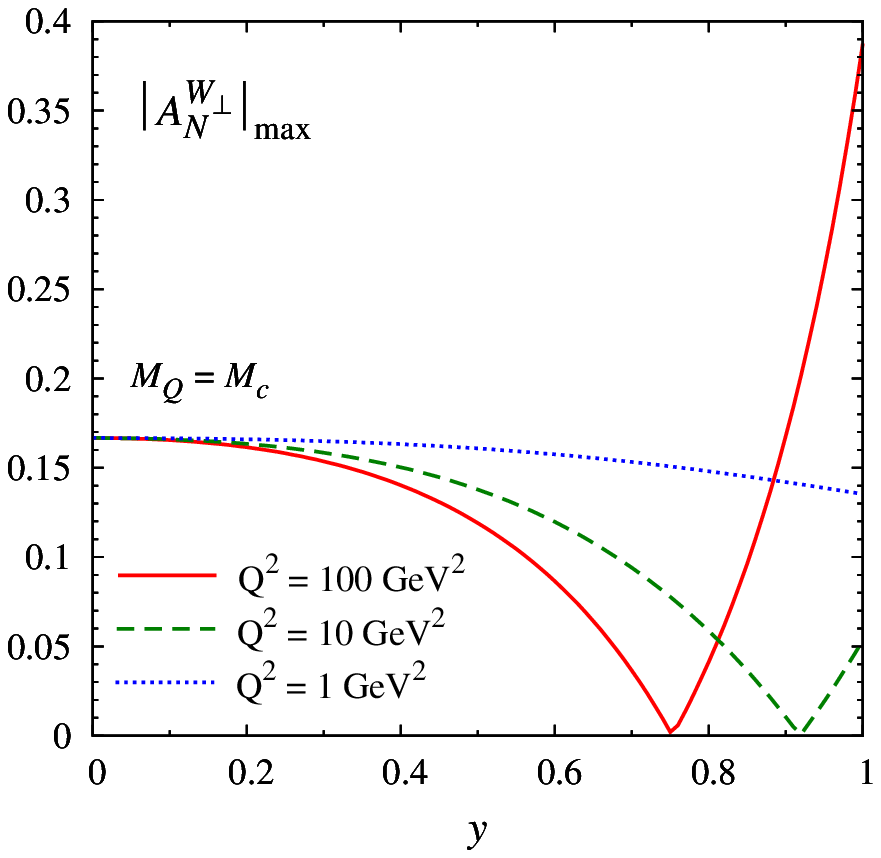}
 \includegraphics[angle=0,width=0.48\textwidth]{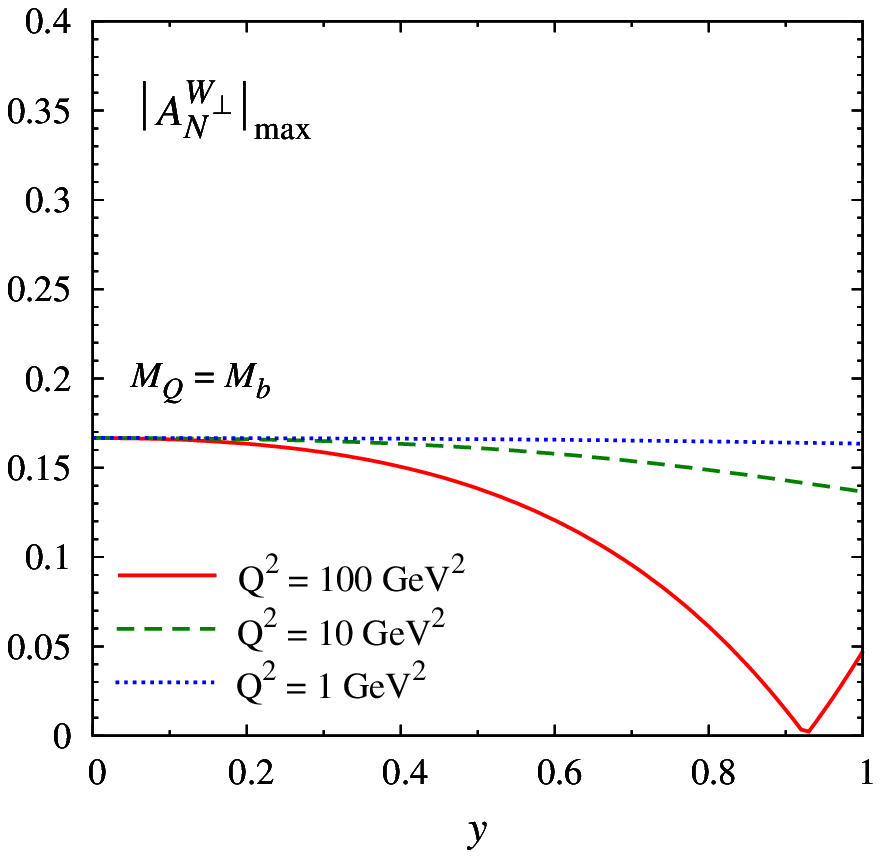}
 \caption{Same as in Fig.~\ref{fig:ANW_y}, but for the upper bounds $\vert A_N^W\vert_{\rm max}$ of the asymmetries $A_N^{\sin(\phi^{\perp}_S+\phi^{\perp}_\sT)}$ and $A_N^{\sin(\phi^{\perp}_S-3\phi^{\perp}_\sT)}$.}
 \label{fig:ANW_y_perp}
\end{center}
\end{figure}
\begin{figure*}[t]
\begin{center}
 \includegraphics[angle=0,width=0.48\textwidth]{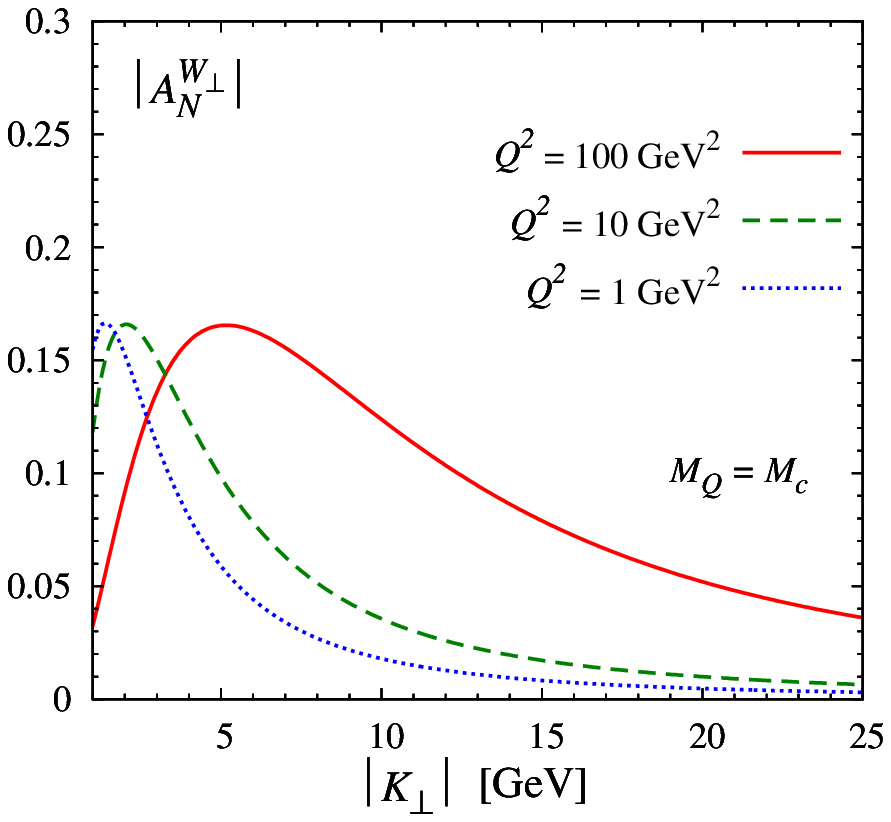}
 \includegraphics[angle=0,width=0.48\textwidth]{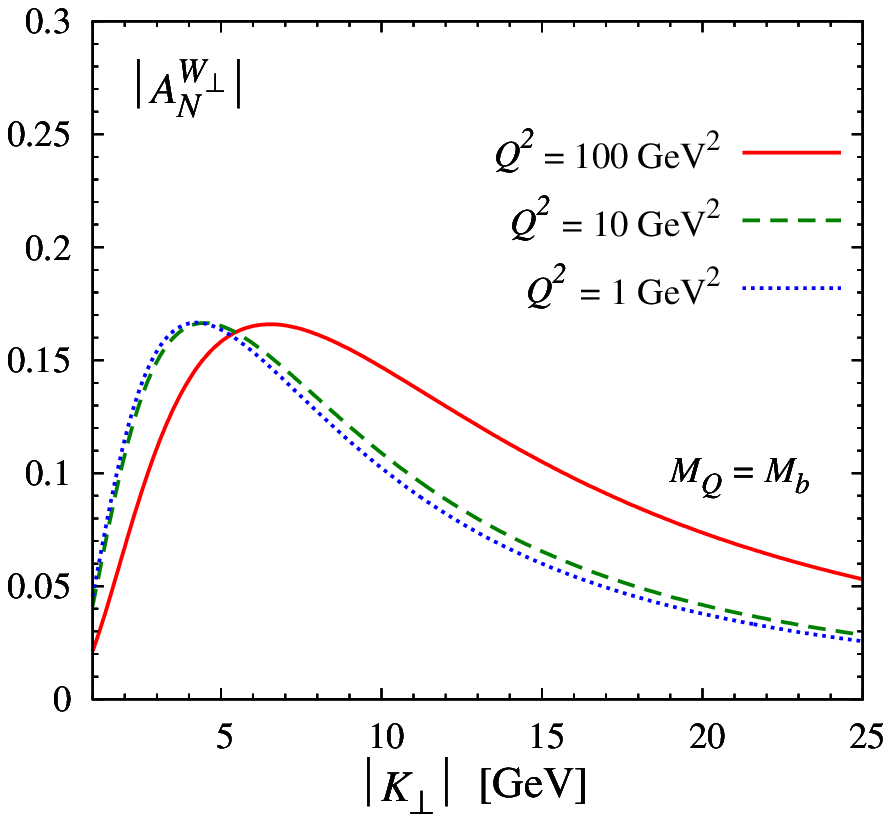}
 \caption{Same as in Fig.~\ref{fig:ANW}, but for the asymmetries $A_N^{W_{\perp}}$, with
 $W_\perp = \sin(\phi^{\perp}_S+\phi^{\perp}_\sT), \,\sin(\phi^{\perp}_S-3\phi^{\perp}_\sT)$.}
 \label{fig:ANWp}
\end{center}
\end{figure*}

\section{\label{sec:links}Sign change test of the T-odd gluon TMDs}

In this section we discuss the gauge link structure of the
gluon TMDs in the process $e\, p^\uparrow \to e' \, Q \bar{Q}\, X$ and its consequences.} 
The subprocess $\gamma^*\, g
\to Q \bar{Q}$ probes a gluon correlator with two future pointing Wilson lines, commonly referred
to as $+$ links, at small $x$ corresponding to the WW type functions. 
In contrast, the process $p^\uparrow\,  p\to \gamma \, \text{jet}\, X$ in the
kinematic regime where gluons in the polarized proton dominate, effectively selects the subprocess
$q \, g \to \gamma \, q$, which probes a gluon correlator with a $+$ and $-$ link (future and past
pointing), at small $x$ corresponding to the dipole type functions. 
These two processes thus probe distinct and entirely independent T-odd gluon TMDs. In
this way processes in $ep$ collisions and $pp$ collisions can be completely complementary. However,
it is important to also note that one can study processes in $pp$ collisions that are fully
related to the ones in $ep$ collisions. For example, the process $p^\uparrow\, p\to \gamma \,
\gamma \, X$ in the back-to-back correlation limit \cite{Qiu:2011ai}, probes T-odd gluon TMDs with
two past-pointing Wilson lines, which are related by an overall sign change to the ones with two
future-pointing links. As a result, we can make the new TMD-formalism prediction that the gluon
Sivers function probed in $e\, p^\uparrow \to e' \, Q \bar{Q}\, X$ is of opposite sign to the one
probed in $p^\uparrow\,  p\to \gamma \, \gamma \, X$:
\begin{equation}
f_{1T}^{\perp\, g \, [e\, p^\uparrow \to e' \, Q \bar{Q}\, X]}(x,p_\sT^2) = - f_{1T}^{\perp\, g \, [p^\uparrow\,  p\to \gamma \, \gamma \, X]}
(x,p_\sT^2),
\end{equation}
and the same relation applies to the other two T-odd gluon TMDs, $h_1^g$ (or $h_{1T}$) and $h_{1T}^\perp$.
Here one can consider instead of a $\gamma \,\gamma$ pair any other color singlet state in
$gg$-dominated kinematics, such as $J/\psi \,\gamma$ or $J/\psi\, J/\psi$
pairs~\cite{Dunnen:2014eta,Lansberg:2015lva}. This is the gluonic analogue of the famous
sign change relation between the quark Sivers TMD probed in SIDIS and
Drell-Yan~\cite{Collins:2002kn}. In our view this provides additional motivation to study gluon
Sivers effects at RHIC or AFTER@LHC and compare it to EIC studies in the future, irrespective of
any theoretical expectations about WW-type gluon TMDs at (very) small $x$. For T-even functions
there is no sign change, such that the process $pp\to  H \, X$ \cite{Sun:2011iw,Boer:2011kf} at LHC
(or instead of the Higgs boson a (pseudo-)scalar heavy quarkonium state \cite{Boer:2012bt}), probes
the same $h_1^{\perp \, g}$ function as the process $e\, p \to e' \, Q \bar{Q}\, X$ at EIC:
\begin{equation}
h_{1}^{\perp\, g \, [e\, p \to e' \, Q \bar{Q}\, X]}(x,p_\sT^2) = h_{1}^{\perp\, g \, [p\,  p\to H \, X]}
(x,p_\sT^2).
\end{equation}
The l.h.s.\ involves the gluon TMDs with two $-$ links and the r.h.s.\ with two $+$ links, which
for T-even functions are equal. On the other hand, a process like Higgs+jet production probes a more complicated link structure~\cite{Boer:2014lka}. The same applies to heavy quark-antiquark pair production in proton-proton collisions or proton-nucleus collisions \cite{Akcakaya:2012si}, although that process is expected to be TMD factorizing at small $x$ only. 

In conclusion, the comparison between TMD observables that can be studied at LHC and at EIC can be related or complementary depending on the process.

\section{Dijet production}
\label{sec:dijet}
The angular structure of the cross section for the process
\begin{equation}
e(\ell) + p(P,S) \to e(\ell^{\prime}) + {\rm jet}(K_1) + {\rm jet}(K_2)+ X\,,
\end{equation}
is the same as in Eqs.~(\ref{eq:cs})-(\ref{eq:csT}), with
\begin{equation}
{\cal N} =   \frac{\alpha^2\alpha_s}{ \pi s \bm K_\perp^2}\, \frac{1}{ \xB y^2}~.
\end{equation}
The terms $A_l^{U/T}$ at LO receive contributions from two subprocesses, $eq \to e^\prime q g$ and
$e g \to e^\prime q \bar q$. They can be written in the form
\begin{align}
A_{l}^U
 & =\sum_{q, \bar q} e_q^2 \,C_F\,{\cal A}_l^{eq\to e q g} \, f_{1}^{q} (x,\bm{q}_{\sT}^2) +\sum_q e_q^2 \,T_R\,{\cal A}_l^{e g\to e q \bar q} \, f_{1}^{g} (x,\bm{q}_{\sT}^2)\, , \nonumber \\
A_{l}^T
 & =\sum_{q, \bar q} e_q^2 \,C_F\,{\cal A}_l^{eq\to e q g} \, f_{1 T}^{\perp\,q} (x,\bm{q}_{\sT}^2) +\sum_q e_q^2 \,T_R\,{\cal A}_l^{e g\to e q \bar q} \, f_{1T}^{\perp\, g} (x,\bm{q}_{\sT}^2) \,,
\end{align}
where $C_F = (N_c^2-1)/2 N_c$, with $N_c$ being the number of colors.
The hard scattering functions ${\cal A}_l^{e q\to e q g}$ are the same as the ones calculated in Ref.~\cite{Pisano:2013cya} for unpolarized scattering. They read
\begin{align}
{\cal A}_0^{e q\to e q g} & =  [1+(1-y)^2]\,{\cal A}_{U+L}^{\gamma^* q\to q g} \,
-y^2 \,{\cal A}_{L}^{\gamma^* q\to q g}\,, \nonumber \\
{\cal A}_1^{e q\to q g} & =  (2-y) \sqrt{1-y}\,{\cal A}_{I}^{\gamma^* q\to q g}\,,\nonumber \\
 {\cal A}_2^{e q\to e q g} & =  2 (1-y)\, {\cal A}_{T}^{\gamma^* q\to q g}\,,\label{eq:Aqstar}
\end{align}
where
\begin{align}
{\cal A}_{U+L}^{\gamma^* q\to q g} & =  \frac{1-z}{ D_0^2}\,
\left\{ 1+z^2 + \left [ 2 z(1-z) + 4z^2(1-z)^2\right ] \,\frac{Q^2}{\bm K_\perp^2} + \left [z^2 (1-z)^2\right] \left [ 1+(1-z)^2\right]\, \frac{Q^4}{\bm K_\perp^4}\right \}\,, \\
{\cal A}_{L}^{\gamma^* q\to q g} & =  4 \,\frac{z^2(1-z)^3}{D_0^2}\,\frac{Q^2}{\bm K_\perp^2} \,,\\
{\cal A}_I^{\gamma^* q\to q g} & =
- 4 \,\frac{z^2(1-z)^2}{D_0^2}\,\left [ 1+ (1-z)^2 \,\frac{Q^2}{\bm K_\perp^2} \right ] \,\frac{Q}{\vert \bm K_\perp \vert}\, ,\\
{\cal A}_{T}^{\gamma^* q\to q g} & =  \,2 \frac{z^2(1-z)^3}{D_0^2}\, \frac{Q^2}{\bm K_\perp^2}\,,
\label{eq:A2ej}
\end{align}
with
\begin{equation}
D_0 \equiv D_0 \left (z, \frac{Q^2}{\bm K_\perp^2} \right ) = 1 + z (1-z) \frac{Q^2}{\bm K_\perp^2}~.
\label{eq:Den0}
\end{equation}
Only the process $e g \to e^\prime q \bar q$  contributes to the terms $B_m^U$,  $B_m^T$ and $B_m^{\prime\,T}$, which can be written
as
\begin{equation}
B_{m}^U  = \sum_{q}e_q^2 \,T_R\,{\cal B}_m^{e g\to e q \bar q} \, \frac{ \bm q_\sT^2  }{M_p^2} \,h_{1}^{\perp\, g} (x,\bm{q}_{\sT}^2)\, ,\qquad B_{m}^T  = B_{m}^U\, \frac{\vert \bm q_\sT \vert }{M_p}\, \frac{h_{1T}^{\perp\, g} (x,\bm{q}_{\sT}^2)}{h_{1}^{\perp\, g} (x,\bm{q}_{\sT}^2)}\, , \qquad B^{\prime\,T}_{m} = B_{m}^U\,\frac{M_p}{\vert \bm q_\sT\vert }\,  \frac{h_{1T}^{g} (x,\bm{q}_{\sT}^2)}{ h_1^{\perp\, g} (x,\bm{q}_{\sT}^2)}~. \nonumber
\end{equation}
The explicit expressions for ${\cal A}_l^{e g\to e q \bar q}$ and ${\cal B}_m^{e g\to e q \bar q}$ can be obtained from the corresponding ones for the process $e g\to e Q  {\overline Q}$, presented in the previous section, by taking the limit $M_Q\to 0$.

For the process under study it is possible to define azimuthal moments similar to those introduced in the previous section for heavy quark pair production. The denominators of the various asymmetries for  dijet production will be
given by
\begin{equation}
A_{0}^U
 =\sum_{q, \bar q} e_q^2 \,C_F\,{\cal A}_0^{eq\to e q g} \, f_{1}^{q} (x,\bm{q}_{\sT}^2) +\sum_q e_q^2 \,T_R\,{\cal A}_0^{e g\to e q \bar q} \, f_{1}^{g} (x,\bm{q}_{\sT}^2)~.
\end{equation}
In a kinematic region where $x$ is small enough, such that the quark contributions can be neglected, the maximal values of the asymmetries for dijet production will be the same as the corresponding ones for heavy quark pair production in the limit $M_Q\to0$. We then recover the expressions of \cite{Dominguez:2011wm,Metz:2011wb} for the linearly polarized gluon asymmetries. In \cite{Dumitru:2015gaa} the effects of the WW $h_1^{\perp \, g}$ in a CGC-formalism calculation including BK evolution with and without running coupling effects were studied for dijet production at EIC. The results for $v_2 \equiv \langle \cos 2 (\phi_\sT-\phi_\perp) \rangle$ were found to be large 
for $\sqrt{s}=100$ GeV, $y=1$, $z=0.5$ and $Q^2= 4 K_\perp^2$, reaching values up to 0.4 at larger transverse momentum of the
dijet pair, close to the bound of 0.5. Although the selected kinematics do not correspond to small $x$ values unless $K_\perp$ is rather small for dijet production considerations ($K_\perp < 5$ GeV in order for $x_B < 0.01$), the theoretically predicted logarithmic suppression \cite{Metz:2011wb} for $p_\sT \ll Q_s$ may thus only become apparent for much smaller $x$ values when $Q_s$ is much larger than typically considered for EIC.

Our MV model estimates for $\vert \langle \cos 2 (\phi_\sT-\phi_\perp) \rangle \vert$ and $\vert \langle \cos 2 \phi_\sT \rangle \vert$ in the dijet case are very similar to the charm production case for $Q^2 \geq 10$ GeV$^2$. For smaller $Q^2$, which corresponds to smaller $x$ values, the asymmetries are unfortunately significantly smaller than the $Q^2=1$ GeV$^2$ curves for $M_Q=M_c$. For this reason we do not include separate plots for those asymmetries. Note that upon inclusion of the quark contributions to the denominator, the asymmetries will decrease further in magnitude.

\begin{figure*}[t]
\begin{center}
\includegraphics[angle=0,width=0.48\textwidth]{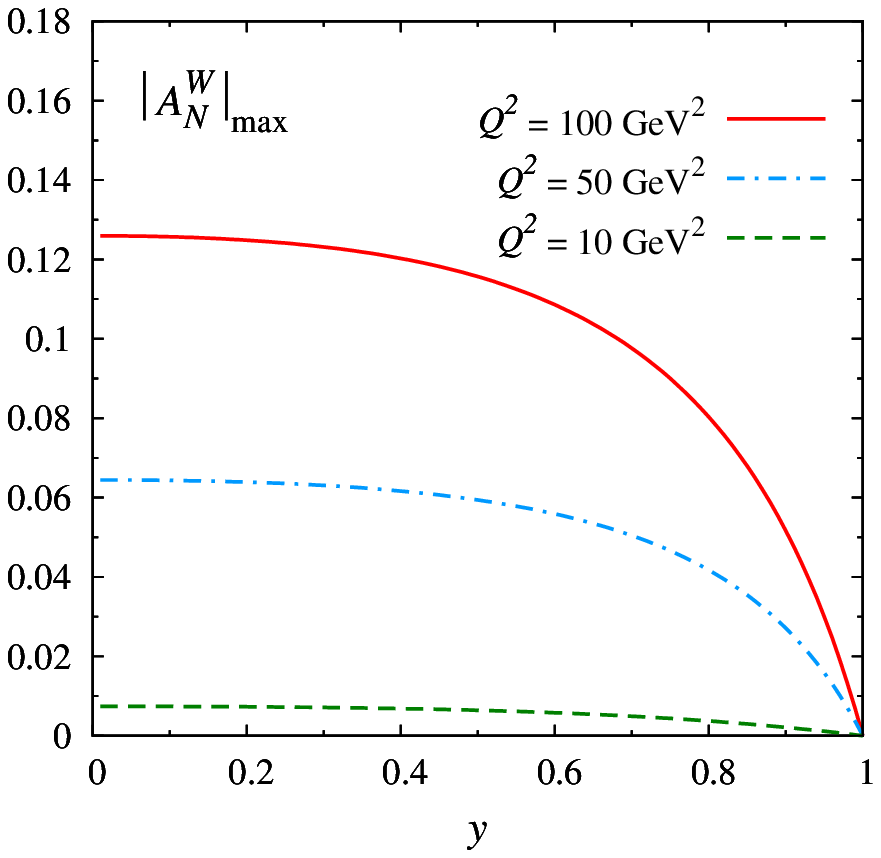}
 \includegraphics[angle=0,width=0.48\textwidth]{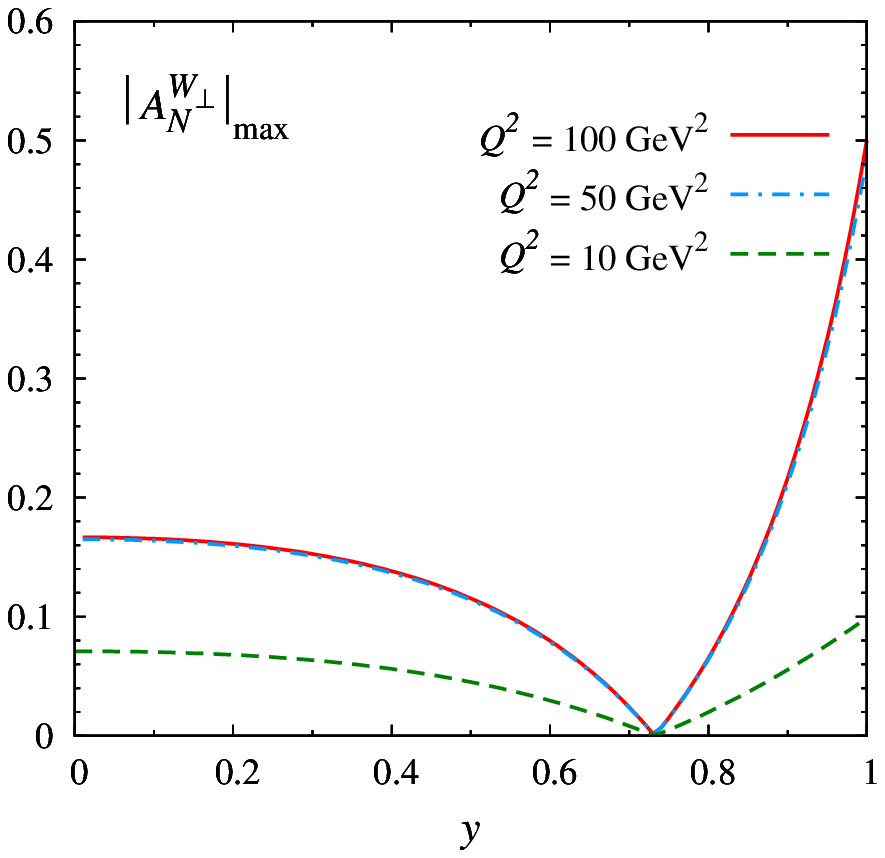}
 \caption{Upper bounds $\vert A_N^W\vert_{\rm max}$, with $W={\sin(\phi_S+\phi_\sT)},\, {\sin(\phi_S-3\phi_\sT)}$  (left panel), and  $\vert A_N^{W_\perp}\vert_{\rm max}$, with $W_\perp ={\sin(\phi^\perp_S+\phi^\perp_\sT)},\, {\sin(\phi^\perp_S-3\phi^\perp_\sT)}$ (right panel),  for the process $e p\to e^\prime {\rm{jet}}\, \rm{jet}\, X$. Estimates are given as a function of $y$, at different values of $K_\perp \ge 4$ GeV  (chosen in such a way that the asymmetries are maximal)  and $Q^2$, with $z=0.5$.}
\label{fig:ANW-y-dijet}
\end{center}
\end{figure*}

\begin{figure*}[t]
\begin{center}
\includegraphics[angle=0,width=0.48\textwidth]{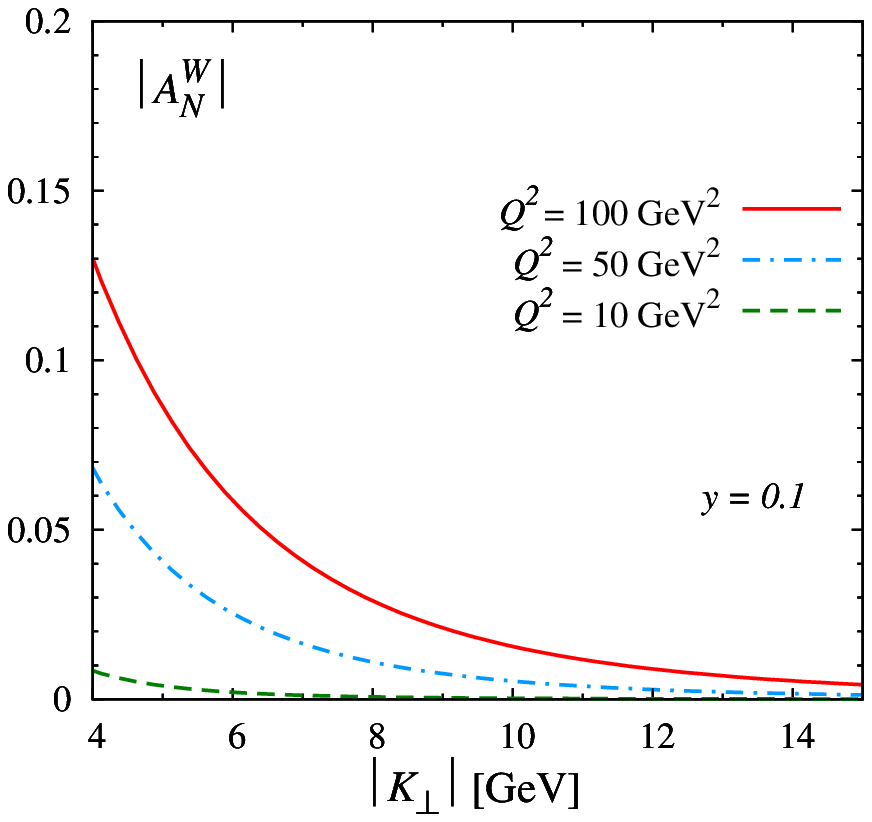}
 \includegraphics[angle=0,width=0.48\textwidth]{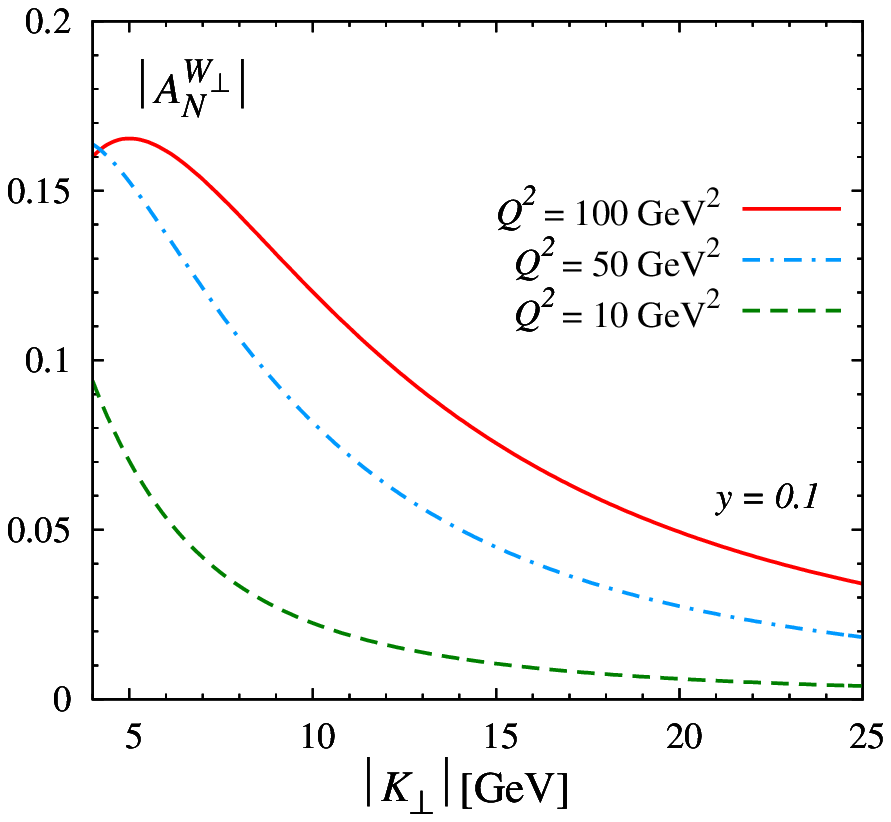}
 \caption{Estimates of $\vert A_N^W\vert $, with $W={\sin(\phi_S+\phi_\sT)}, \,{\sin(\phi_S-3\phi_\sT)}$ (left panel), and $\vert A_N^{W_\perp}\vert $, with $W_\perp ={\sin(\phi^\perp_S+\phi^\perp_\sT)},\, {\sin(\phi^\perp_S-3\phi^\perp_\sT)} $ (right panel),  for the process $e p\to e^\prime {\rm{jet}}\, \rm{jet}\, X$. The asymmetries are calculated at $z=0.5$,  $y=0.1$ and at different values of  $Q^2$, as a function of $\vert \boldsymbol K_\perp\vert$ ($\ge$ 4 GeV).}
\label{fig:ANW-dijet}
\end{center}
\end{figure*}

For the T-odd gluon TMD asymmetries, upper bounds (without quark contributions) are shown in Figs.~\ref{fig:ANW-y-dijet} and \ref{fig:ANW-dijet}. Again they are similar to the $M_Q=M_c$ case for large $Q^2$ and substantially smaller for small $Q^2$. We emphasize that at small $x$ these bounds are not expected to be satisfied. The simple models presented 
could be used to scale the asymmetries to more reasonable values in the small-$x$ case. We have selected $K_\perp \ge 4$ GeV based on
the multijet analysis of small $x_B$ HERA data in \cite{Chekanov:2007dx}.  


\section{Conclusions}
\label{sec:concl}

We have studied the azimuthal asymmetries in heavy quark pair and dijet production in DIS process,
which provide direct access to the WW-type gluon TMDs. We found that in these processes 
the gluon TMDs arise in the same way as the quark TMDs do in semi-inclusive DIS. The measurements 
of these azimuthal asymmetries at an EIC would allow a sign change test of the T-odd gluon TMDs 
by comparing to corresponding observables at RHIC and AFTER$@$LHC. Furthermore, it was shown 
that the asymmetries maximally allowed by the positivity bounds are rather sizeable, except in some kinematic limits. 
Although WW-type gluon TMDs suffer from suppression in the small-$x$ region, the effects from linearly polarized 
gluons are still expected to be sizeable in heavy quark pair production at an EIC, but less so in dijet production 
because of the requirement that the transverse momentum of each jet must be large. 
For the Sivers gluon TMD the bounds are always maximal and provide little guidance, especially in the small-$x$ region 
where the WW-type function is suppressed by a factor of $x$ with respect to the unpolarized gluon TMD. 
For the other two T-odd gluon TMDs that arise with $\sin(\phi_S+\phi_\sT)$ and $\sin(\phi_S-3\phi_\sT)$ modulations, 
their ratio can be exploited to test small-$x$ expectations. We have provided some simple model expectations for that ratio. 
It will also be interesting to compare these quantities to the dipole type T-odd gluon TMDs that arise in other processes, such as 
virtual photon plus jet production in polarized proton-proton collisions at RHIC. This 
allows to experimentally test the theoretical expectation that the WW and dipole type T-odd gluon TMDs have 
different small $x$ asymptotic behavior. Hence, the studies of these asymmetries could form a prominent part of 
both the spin physics program and the small-$x$ physics program at a future EIC.

\acknowledgments

This research is partially supported by the European Research Council (ERC) under the FP7 ``Ideas'' programme (grant agreement No.~320389, QWORK) and the European Union's Horizon 2020 research and innovation programme (grant agreement No. 647981, 3DSPIN).

\end{document}